\documentclass[useAMS,usenatbib]{mn2e}
\usepackage{graphics}
\usepackage{supertabular}

\usepackage{tabularx}

\usepackage{epic}

\usepackage{rotating}

\usepackage{xcolor}

\usepackage{subfigure}

\usepackage{caption}

\usepackage{amsmath}
\usepackage{amssymb}

\usepackage{aas_macros}

\usepackage{pdflscape}

\newcommand{\tcm}{21cm}

\title[Erasing the Milky Way]{Erasing the Milky Way: new cleaning technique applied to GBT intensity mapping data}
\author[L. Wolz et al.]{L. Wolz$^{1, 2}$\thanks{E-mail:lwolz@unimelb.edu.au}, C. Blake$^3$, F. B. Abdalla$^4$, C.J. Anderson$^5$, 
 T.-C. Chang$^6$, \newauthor  Y.-C. Li$^7$, K.W. Masui$^{8,9}$, E. Switzer$^{10}$, U.-L. Pen$^{11}$,  T.C. Voytek$^{12}$, J. Yadav$^{13}$ \\
$^1$School of Physics, University of Melbourne, Parkville, VIC 3010, Australia\\
$^2$ARC Centre of Excellence for All-Sky Astrophysics (CAASTRO)\\
$^3$Centre for Astrophysics \& Supercomputing, Swinburne University of Technology, P.O. Box 218, Hawthorn, VIC 3122, Australia\\
$^4$Department of Physics and Astronomy, University College London, London WC1E6BT, UK\\
$^5$Department of Physics, University of Wisconsin - Madison, Madison, WI, USA\\
$^6$Academia Sinica Institute of Astronomy and Astrophysics, P.O. Box 23-141,Taipei, 10617 Taiwan \\
$^7$National Astronomical Observatories, Chinese Academy of Sciences, 20A Datun Road, Chaoyang District, Beijing 100012, China\\
$^8$Department of Physics and Astronomy, University of British Columbia, 6224 Agricultural Rd., Vancouver, V6T 1Z1, Canada\\
$^9$Canadian Institute for Advanced Research, CIFAR Program in Cosmology and Gravity, Toronto, ON, M5G 1Z8, Canada\\
$^{10}$NASA Goddard Space Flight Center, Greenbelt, MD 20771, USA\\
$^{11}$Canadian Institute for Theoretical Astrophysics, 60 St. George Street, Toronto, ON, M5S 3H8, Canada\\
$^{12}$Astrophysics and Cosmology Research Unit, School of Chemistry and Physics, University of KwaZulu-Natal, Durban, 4041, South Africa\\
$^{13}$Sri Aurobindo College, University of Delhi, Malviya Nagar, New Delhi 110017, India }
\begin{document}

\date{}

\pagerange{\pageref{firstpage}--\pageref{lastpage}} \pubyear{}

\maketitle

\label{firstpage}

\begin{abstract}

We present the first application of a new foreground removal pipeline
to the current leading HI intensity mapping dataset, obtained by the
Green Bank Telescope (GBT).  We study the 15hr and 1hr field data of
the GBT observations previously presented in
\cite{2013ApJ...763L..20M} and \cite{2013MNRAS.434L..46S}, covering
about 41 square degrees at $0.6<z<1.0$, for which cross-correlations
may be measured with the galaxy distribution of the WiggleZ Dark
Energy Survey.  In the presented pipeline, we subtract the Galactic
foreground continuum and the point source contamination using an
independent component analysis technique (\textsc{fastica}), and
develop a Fourier-based optimal estimator to compute the temperature
power spectrum of the intensity maps and cross-correlation with the
galaxy survey data.  We show that \textsc{fastica} is a reliable tool
to subtract diffuse and point-source emission through the non-Gaussian
nature of their probability distributions. {The
  temperature power spectra of the intensity maps is dominated by
  instrumental noise on small scales which \textsc{fastica}, as a
  conservative subtraction technique of non-Gaussian signals, can not
  mitigate.}  However, we determine similar GBT-WiggleZ
cross-correlation measurements to those obtained by the Singular Value
Decomposition (SVD) method, and confirm that foreground subtraction
with \textsc{fastica} is robust against 21cm signal loss, as seen by
the converged amplitude of these cross-correlation measurements.
{We conclude that SVD and \textsc{fastica} are
  complementary methods to investigate the foregrounds and noise
  systematics present in intensity mapping datasets.}

\end{abstract}


\section{Introduction}

Cosmological observations aim to map the largest possible volume of
the Universe in order to develop a better understanding of the
formation and evolution of large-scale structure.  The clustering of
galaxies traces both major unknown ingredients of the standard model
of cosmology: the dark matter distribution, and thus the laws of
gravity, in addition to the time-dependent expansion of the Universe
driven by dark energy.  Historically optical galaxy surveys, such as
Sloan Digital Sky Survey \citep{Tegmark:2003uf} or the WiggleZ Dark
Energy Survey \citep{2010MNRAS.401.1429D}, have been used to map
large-scale structure by cataloguing the angular positions and
redshifts of galaxies.  While achieving major scientific discoveries
such as the detection of the Baryon Acoustic Oscillations (BAO)
\citep{Eisenstein:2005su, Percival:2007yw, 2010MNRAS.401.2148P}, this
approach is affected by, for instance, selection effects and redshift
inaccuracies for photometric surveys, and low survey speed for
spectroscopic observations.

Recent advances in radio interferometry, both instrumental and
algorithmic, have created excellent prospects for forthcoming radio
surveys to efficiently map large-scale structure.  In addition to the
traditional galaxy surveys which find sources above a flux threshold
in radio data cubes, a new observational method called intensity
mapping has been formulated as described by
e.g. \cite{2004MNRAS.355.1339B,2009atnf.prop.2491V,Peterson:2012hb,2014arXiv1405.1452B}.
This technique exploits the low angular resolution of radio telescopes
by efficiently mapping the integrated spectral line emission with a
beam small enough to resolve the BAO scale. The neutral hydrogen line
(HI) at \tcm is an excellent tracer of the galaxy distribution and not
prone to line confusion \citep{2011ApJ...740L..20G}.  Intensity
mapping has also been envisaged using different spectral lines such as
the rotational CO lines \citep{2011ApJ...741...70L} or the Lyman-alpha
line \citep{2013arXiv1309.2295P}.

In comparison with galaxy surveys, intensity mapping has the advantage
of measuring the entire HI flux in the observed frequency channel.
This implies that there are no observational selection effects, which
allows access to a wide redshift range, and the integrated luminosity
function is probed rather than the most luminous objects.  The
challenges of intensity mapping are the demands on instrumental
stability and the high Galactic foregrounds which dominate the
targeted frequency ranges.

The challenge of Galactic foreground subtraction has been extensively
addressed in the framework of Cosmic Microwave Background (CMB)
observations, see \cite{Adam:2015wua} and \cite{ Ade:2015qkp} for the
latest results.  The foregrounds for intensity mapping have fewer
existing observational constraints than the microwave sky, however,
the data contains more line-of-sight information as it extends over a
wider frequency interval.  Most foreground separation methods utilise
the power-law dependence of the foregrounds in the frequency
direction, where the techniques can be divided into parametric methods
\citep{2008arXiv0807.3614A,Shaw2013,2015PhRvD..91h3514S,
  2016ApJS..222....3Z} and blind methods \citep{Wolz:2013wna,
  Switzer:2015ria, 2016MNRAS.456.2749O}.  In this work, we perform
foreground subtraction using an Independent Component Analysis
(\textsc{fastica}, \citealt{DBLP:journals/tnn/Hyvarinen99}) motivated
by its previous successful applications to CMB
{simulations} \citep{Maino:2001vz}, epoch of
reionization studies \citep{Chapman:2012yj} and intensity mapping
simulations \citep{Wolz:2013wna}.

After promising theoretical predictions of intensity mapping surveys
\citep{Wyithe:2007rq,Chang:2007xk}, the Green Bank Telescope (GBT)
team has pioneered the realisation of an experiment and data analysis
as shown by \cite{Chang:2010jp}, \cite{2013ApJ...763L..20M} (MA13
hereafter) and \cite{2013MNRAS.434L..46S} (SW13 hereafter).  The
foreground removal presented by SW13 is based on a singular value
decomposition (SVD) where the highest eigenvalues assumed to
contain the foregrounds are subtracted from the data.  MA13 show
the detection of the intensity mapping signal in cross-correlation
with the WiggleZ Dark Energy Survey, which is used by SW13 in
combination with the auto-power spectrum to constrain the amplitude of
the correlation as $\Omega_{\rm HI}b_{\rm HI} = [0.62\pm 0.23]\times
10^{-3}$, where $\Omega_{\rm HI}$ is the neutral hydrogen energy
density and $b_{\rm HI}$ is the HI bias parameter.

In this work, we apply our foreground removal and power spectrum
estimator pipeline to the GBT datasets and demonstrate how
\textsc{fastica} can reliably subtract foregrounds, in addition to
providing insight into the signal properties.

This paper is structured as follows: Sec.\ref{obs} briefly outlines
the data specifications of the GBT observations and the WiggleZ Dark
Energy Survey.  Sec.\ref{PS} describes the Fourier-based power
spectrum estimator for the auto- and cross-correlations.  Sec.\ref{FG}
presents a detailed analysis of the component separation and the data
properties as revealed by this analysis.  In Sec.\ref{res}, the
intensity mapping power spectrum and the cross-correlation with
WiggleZ are determined and compared with the previous GBT results.  We
conclude in Sec.\ref{conc}.

\section{Observations}
\label{obs}

\subsection{Green Bank Telescope Intensity Maps}

A detailed description of the observing strategy of the GBT intensity
maps can be found in MA13, and we provide a short summary.  The
intensity maps we analyze consist of a $4.5 \times 2.4\deg$ ``15hr
deep field'' centered at RA=14h31m28.5s and Dec=$2\deg0'$, which was
observed with 105h integration time, and a $7.0 \times 4.3\deg$ ``1hr
wide field'' centered at RA=0h52m0s and Dec=$0\deg9'$, which was
observed with 85h integration time.  Each of the fields was observed
in 4 sub-datasets $\{\rm A, B, C, D\}$ which have similar integration
time and sky coverage. The subset maps were taken at different times,
such that the thermal noise of the instrument is independent in each
map.

The data were obtained in the frequency range 700-900 MHz,
i.e. $0.58<z<1$ for the redshifted 21cm line, divided into 4096
channels across the bandwidth.  The data was rebinned into frequency
bands of width 0.78 MHz, equivalent to $3.8h^{-1}{\rm Mpc}$ comoving
width along the line of sight at the band center.  The total
calibration uncertainty is 9\%.  The map-making conventions of the GBT
team follow the CMB description given by \cite{1997ApJ...480L..87T}.
The angular pixels have dimension $0.0627 \times 0.0627\deg$ and the
maps consist of $78 \times 43$ pixels for the 15hr field and $161
\times 83$ pixels for the 1hr field.  {The telescope
  beam has a co-moving width of approximately $9.6h^{-1}{\rm Mpc}$ at
  the band center, corresponding to the Full-With at Half-Maximum (FWHM)
   $\theta_{\rm FWHM}=0.28\deg$ of the symmetric, two-dimensional Gaussian shaped telescope beam.  In
  the analysis presented by MA13 and SW13, the data is convolved to a
  common angular resolution $\theta_{\rm FWHM}=0.44\deg$ to mitigate
  the effects of polarization leakage.  In this work we instead
  process the unconvolved data with a frequency-dependent resolution
  spanning $0.25<\theta_{\rm FWHM}<0.31\deg$ across the observed
  range.  The telescope beam can be well-approximated by a Gaussian
  with standard deviation $\theta_{\rm FWHM}/2$.}

\subsection{WiggleZ Dark Energy Survey}

The WiggleZ Dark Energy Survey \citep{2010MNRAS.401.1429D} is a
large-scale galaxy redshift survey of bright emission-line galaxies
over the redshift range $z < 1$, with median redshift $z \approx 0.6$
and galaxy bias factor $b \sim 1$.  The survey was carried out at the
Anglo-Australian Telescope between August 2006 and January 2011.  In
total $\sim 200{,}000$ redshifts were obtained, covering $\sim 1000$
deg$^2$ of equatorial sky divided into seven well-separated regions.
The two GBT fields analyzed in this study have nearly complete angular
and redshift overlap with two of these WiggleZ regions, and the two
datasets are therefore well-suited for cross-correlation analysis.
Following the cut to the redshift range $0.58 < z < 1$, a total of
6731 WiggleZ galaxies are used in this analysis.  The WiggleZ
selection function within each region, which is used to produce the
optimal weighting for our power spectrum analysis, was determined
using the methods described by \cite{2010MNRAS.406..803B}, averaging
over a large number of random realizations matching the angular
completeness and redshift distribution of the sample.

\section{Power spectrum measurement}
\label{PS}

\subsection{Optimally-weighted power spectrum estimator}

The sky area and redshift interval of the GBT intensity mapping data
allows us to apply a ``flat-sky approximation'' where we map the
angular and redshift pixels into a cuboid in comoving space using a
fiducial cosmology.  Our description in this section follows the
conventions of \cite{2010MNRAS.406..803B} and
\cite{2013MNRAS.436.3089B}, and recasts the analysis in terms of
temperature power spectra of intensity maps with a weighting scheme
dictated by the noise properties of the observations.

We consider the intensity maps as overtemperatures in units of mK
measured as a discrete function of position, $\delta(\vec x_i) =
T(\vec x_i) - \bar T$, where $\bar T$ is the mean temperature of each
frequency slice.  The pixel dimensions of the data are
$(N_x,M_y,K_z)$, where $N_x$ and $M_y$ define the angular grid given
by the map-making process and $K_z$ is the total number of frequency
bins.  The total number of pixels is $N_{pix} = N_x \cdot M_y \cdot
K_z$.  The data cuboid has co-moving physical dimensions $L_x \times
L_y$ on the sky, and radial dimension $L_z$, where we neglect the slow
variation of co-moving pixel size with frequency such that each cell
has volume $V_{\rm cell} = \frac{L_x \cdot L_y \cdot L_z}{N_x M_y
  K_z}$.  We use a fiducial cosmological model given
{by \cite{2015arXiv150201589P} with parameters $\vec
  \theta =(h=0.678, \Omega _{\rm m}=0.308, \Omega _{\rm b}=0.0486,
  n_{\rm s}=0.968, \sigma_{8}=0.816, w=-1.0)$.}

The Fourier-transformed temperature field is a function of wavevector
$\vec k_l$.  The resolution of the measurements in each direction of
Fourier-space is given by $\Delta k_x=2\pi/L_x$, $\Delta k_y=2\pi/L_y$
and $\Delta k_z=2\pi/L_z$.  The upper bound on $\vec k_l$, which
refers to the smallest scale in real space which can be measured in
our grid, is determined by the Nyquist frequency in each direction
$k_{\rm{Nyq},x}=\pi\cdot N_x/L_x$, $k_{\rm{Nyq},y}=\pi\cdot M_y/L_y$
and $k_{\rm{Nyq},z}=\pi\cdot K_z/L_z$.  The Fourier amplitudes for
each mode are calculated via
\begin{equation}
 \tilde \delta(\vec k_l)=\sum_{j=1}^{N_{pix}} \delta(\vec x_j) w(\vec
 x_j) \exp{(i \vec k_l \cdot \vec x_j)}.
\end{equation}
The temperature of each pixel is multiplied by a weighting function
$w(\vec x_j)$, which we normalise such that $\sum_{i=1}^{N_{pix}}w(\vec
x_i)=1$ in the estimators given below.

In the case of noise-dominated intensity mapping data, the weighting
function is directly related to the noise in each pixel. We consider a
simple inverse-variance weighting using this noise map.  Under the
assumption that the noise is uncorrelated between pixels, the estimate
of the power spectrum for each Fourier amplitude in volume units is
\begin{equation}
P_{\rm{est}}(\vec k_l)=\frac{V_{\rm cell}|\tilde \delta(\vec k_l)|^2
}{\sum_{j=1}^{N_{pix}}w^2(\vec x_j)}
\label{3dPS}
\end{equation}
In our analysis, we estimate the cross-power spectrum of every pair of
different sub-dataset maps, in order to suppress the additive thermal
noise correction term.  The cross-power spectrum for two intensity
mapping datasets $A$ and $B$ is
\begin{equation}
P^{AB}_{\rm{est}}(\vec k_l)=\frac{V_{\rm cell} \mathrm{Re}\{ \tilde
  \delta^A(\vec k_l)\cdot \tilde\delta^B(\vec k_l)^*\}
}{\sum_{j=1}^{N_{pix}}w^A(\vec x_j)\cdot w^B(\vec x_j)}
\label{3dPScross}
\end{equation}
We bin amplitudes of Fourier modes $\vec k$ according to the
value of $k=|\vec k|$.

The above equation for the cross-correlation between two intensity
mapping datasets can be recast for the cross-correlation with galaxy
survey data, $P^{\rm X}_{\rm{est}}(\vec k_l)$, where the overdensity
is defined as the number of galaxies per voxel $N_i$ divided by the
mean galaxy density at this position of the cube $\bar N(\vec x_i)$,
$\delta_{\rm g}(\vec x_i)=N_i/\bar N(\vec x_i)$. The optimal weighting
function $w_{\rm g}(\vec x_i)=1/(1+W(\vec x_i)\times \bar N P_0)$ is
computed via the selection function $W(\vec x_i)$ given by
\citep{2010MNRAS.406..803B} with $P_0=10^3 h^{-3}\rm{Mpc}^3$.

{We also correct the power spectrum estimate for the
  effect of the telescope beam by dividing the measured power spectrum
  $\hat P(\vec k_i)$ by the discretized, Fourier-transformed beam
  $\tilde B(\vec k_i)$.  The beam $B(\vec x_j)$ is constructed as a
  spatial, 2-dimensional Gaussian discretized on the grid such that it
  only acts on modes perpendicular to the line-of-sight.}

The thermal noise contributes to errors in the cross-power spectrum
measurements.  For noise-dominated data, the cosmic variance
contribution can be neglected.  Under the assumption that the noise
has similar properties in each dataset, we can estimate the error in
the intensity mapping cross-correlation as (compare
e.g. \citealt{2009MNRAS.397.1348W})
\begin{equation}
  \sigma(P^{AB}_{\rm{est}}(k_i))=P_{\rm noise}(k_i)/\sqrt{2\cdot N(k_i)}
\label{autocorrnoise}
\end{equation}
where the noise power spectrum $P_{\rm noise}$ is scaled by $\sqrt{2}$
since two independent maps are correlated, and $N(k_l)$ is the number
of independent measured modes per bin.  There are various approaches
for estimating $P_{\rm noise}$, which we discuss further in
Sec.~\ref{res}.

The error in the galaxy-temperature cross-power spectrum can be
estimated using the galaxy power spectrum $P^{\rm g}_{\rm{est}}$ and
the intensity mapping power spectrum $P^{AB}_{\rm{est}}$
\begin{equation}
  \sigma(P^{\rm X}_{\rm{est}}(k_i)) = \sqrt{\frac{1}{2\cdot N(k_i)}}
  \sqrt{P^{\rm X}_{\rm{est}}(k_i)^2 + P^{\rm g}_{\rm{est}}(k_i)
    P^{AB}_{\rm{est}}(k_i)}
\label{Xnoise}
 \end{equation}
In this work we present all power spectra in the dimensionless form
\begin{equation}
\Delta^2(k_i)=\frac{k_i^3}{2\pi^2}P(k_i).
\end{equation}

\subsection{Theoretical prediction}
\label{PSth}

We compare the measured power spectra of the intensity maps to a
theoretical prediction $P_{\rm th}(k)$, generated from the linear CAMB
(\citealt{Lewis:1999bs}) power spectrum scaled by the growth function
for $z=0.8$.  The weighting scheme alters the shape of the power
spectrum.  In order to account for this effect, we convolve the
theoretical prediction with the weighting function via
\begin{equation}
 \hat P_{\rm th}(\vec k_j)=\frac{\sum_{i} P_{\rm th}(\vec k_i^\prime)
   \mathrm{Re}\{\tilde w_A(\vec k_j-\vec k_i^\prime) \tilde w_B(\vec
   k_j-\vec k_i^\prime)^* \}}{\sum_{j=1}^{N_{pix}}w^A(\vec
   x_j)w^B(\vec x_j) }.
\end{equation}
For this computation we grid the 1-dimensional $P_{\rm th}(k)$ in 3D
Fourier space in the same fashion as the intensity maps, hence
discretizing the modes as $P_{\rm th}(\vec k_j)$.

The estimated power spectrum of the intensity maps relates to the
theory as $P_{\rm{est}} = b_{\rm HI}^2 \bar T_{\rm HI}^2 P_{\rm th}$.
We use Equ. 1 in MA13 as a model for the mean HI temperature, which
predicts $\bar T_{\rm HI} = 0.29\rm{mK} \times \Omega_{\rm HI}
/10^{-3}$ in our fiducial cosmology.  The factor $b_{\rm
  HI}\Omega_{\rm HI}$ is chosen following MA13 as $0.43\times
10^{-3}$.  We note that this is a lower limit, because of the unknown
cross-correlation coefficient $r$ between HI and galaxy overdensity.
The cross-correlation $P^{\rm X}_{\rm{est}} = b_{\rm HI} b_{\rm opt}
\Omega_{\rm HI} r P_{\rm th}$ depends additionally on the optical
galaxy bias, which is assumed to be $b^2_{\rm opt} = 1.48$ according
to the measurements in \cite{2010MNRAS.406..803B}.

\section{Foreground removal and systematics analysis}
\label{FG}

\subsection{\textsc{Fastica} application}

We apply \textsc{fastica} to the intensity mapping data cube in order
to remove the foregrounds.  We refer the reader to \cite{Wolz:2013wna}
for a more detailed description of the method, and provide a brief
summary here.  The methodology solves the linear problem
\begin{equation}
\boldsymbol x= \mathbf A \boldsymbol s + \epsilon=
\sum_{i=1}^{N_{\rm{IC}}}\boldsymbol{a_i} s_i + \epsilon,
\label{eq:ica}
\end{equation}
where $\boldsymbol x$ is the input data, $\mathbf A$ is a mixing
matrix, $\boldsymbol s$ represents the $N_{\rm IC}$ independent
component amplitudes (ICs), and $\epsilon$ is the residual.  The ICs
can be interpreted as maps with the same spatial dimension as the
intensity maps.  The amplitude of each IC as a function of frequency
is given by the mixing modes $\boldsymbol{a_i}$.  \textsc{Fastica}
identifies components with strong spectral correlation and
incorporates them into the ICs by using the Central Limit theorem,
such that the non-Gaussianity of the probability density function of
each IC is maximized.  This implies that \textsc{fastica} neglects
Gaussian-distributed components, such that the contributions
represented by $\mathbf A \boldsymbol s$ include
{non-Gaussian foregrounds (and potentially non-Gaussian
  HI signal and noise)}.  The residual $\epsilon$ is the
foreground-subtracted data cube, which ideally contains Gaussian 21cm
signal and noise, but in principle can also include residual
foregrounds.

The number of ICs ($N_{\rm IC}$) used in the component separation is a
free parameter and can not be determined by \textsc{fastica} itself.
In the following sub-sections we carefully examine the
foreground-subtracted data for different numbers of ICs, ensuring that
the results do not sensitively depend on this choice.

\subsection{Foreground point source removal}

\begin{figure*}
\subfigure[2 ICs]{
\includegraphics[width=0.5\textwidth, clip=true, trim=0 40 0 40]{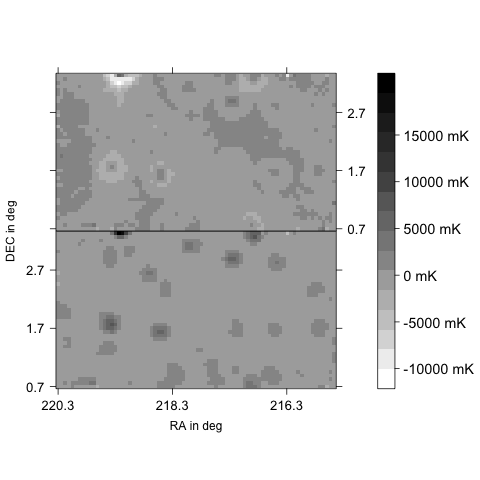}}\subfigure[2 ICs; masked]{
\includegraphics[width=0.5\textwidth, clip=true, trim=0 40 0 40]{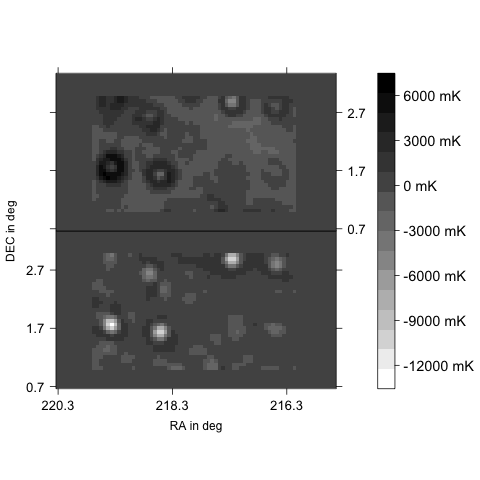}}
\subfigure[6 ICs]{
 \includegraphics[width=0.5\textwidth, clip=true, trim=0 60 0 60]
{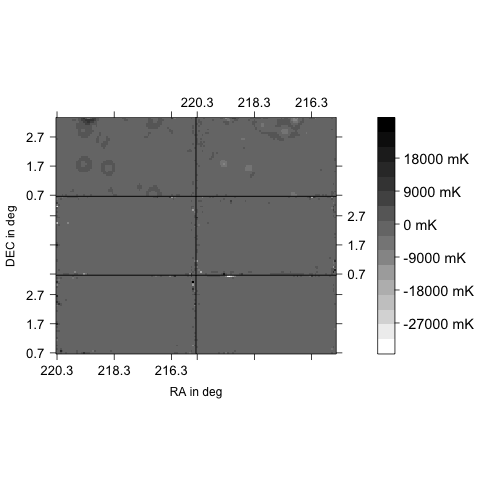}}\subfigure[6 ICs; masked]{
\includegraphics[width=0.5\textwidth, clip=true, trim=0 60 0 60]
{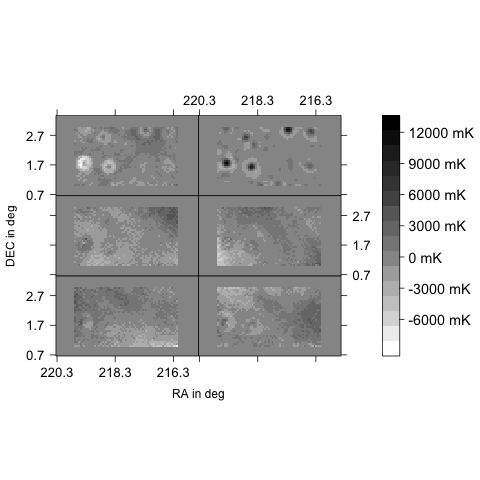}}
\subfigure[10 ICs]{
\includegraphics[width=0.5\textwidth, clip=true, trim= 0 100 0 100]
{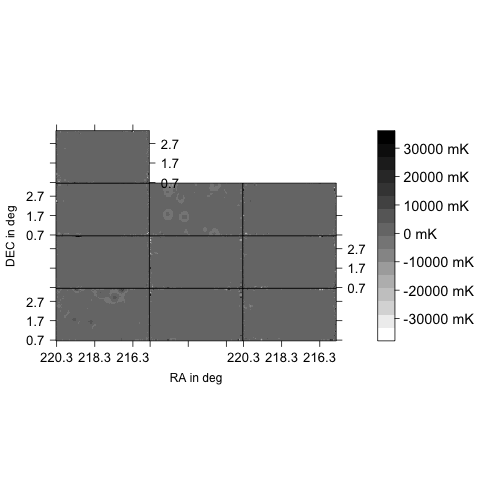}}\subfigure[10 ICs; masked]{
\includegraphics[width=0.5\textwidth, clip=true, trim= 0 100 0 100]
{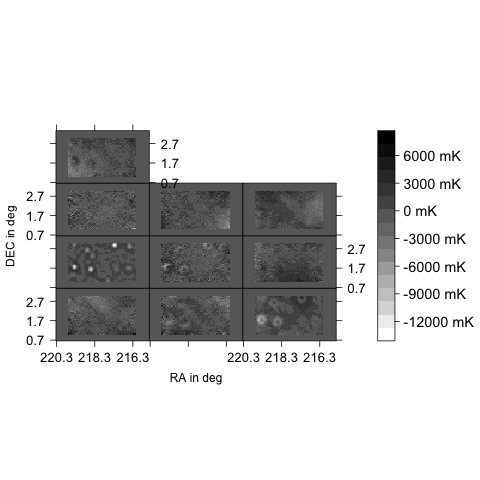}}
\caption{The maps of the independent components identified for
  sub-dataset A of the 15hr-field, assuming different numbers of ICs.
  The first column shows the results when analyzing the full field,
  and the second column displays a masked analysis in which the
  noise-dominated edges are disregarded.}
\label{ICs}
\end{figure*}

The data maps contain prominent signals from extra-Galactic point
sources which contribute emission at all frequencies.
\textsc{Fastica} models the spatial structure of the foregrounds as
well as their frequency dependence.  Fig.~\ref{ICs} presents the maps
of the ICs for the sub-dataset A of the 15hr-field, where an analysis
with 2, 6 and 10 ICs is shown.  The first column displays the IC maps
determined by an analysis of the full field.  In these maps, the IC
model is dominated by features at the edges of the fields driven by
high instrumental noise in these regions, due to the poor
observational coverage of the edges of the fields.  The IC maps do not
optimally model the point source structure and diffuse foregrounds
because of this high noise contamination.  By masking out those
regions, as seen in the second column of Fig.~\ref{ICs}, the ICs
instead contain the spatial structure of the point sources of
{extra-Galactic} foregrounds. We observe similar
behaviour of the ICs for the remaining datasets and the 1hr-field
analysis, hence we will use the masked data cubes for our analysis.

\begin{figure*}
\subfigure[2 ICs]{
\includegraphics[width=0.5\textwidth, clip=true, trim= 0 100 0 130]
{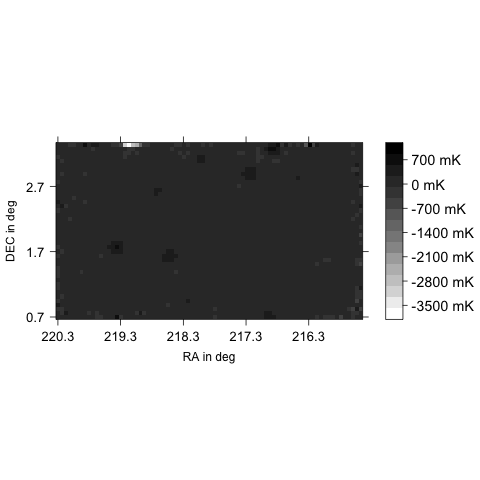}}\subfigure[2 ICs; masked]{
\includegraphics[width=0.5\textwidth, clip=true, trim= 0 100 0 130]
{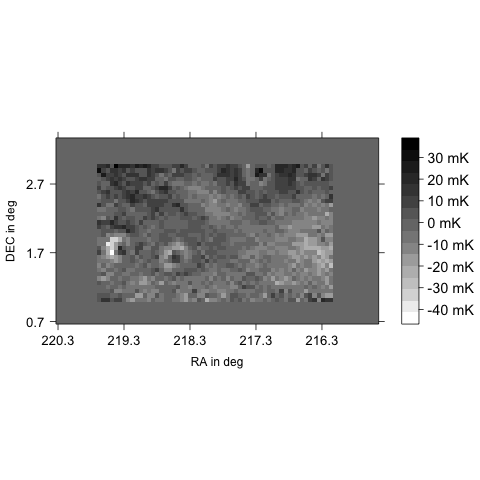}}
\subfigure[6 ICs]{
\includegraphics[width=0.5\textwidth, clip=true, trim= 0 100 0 130]
{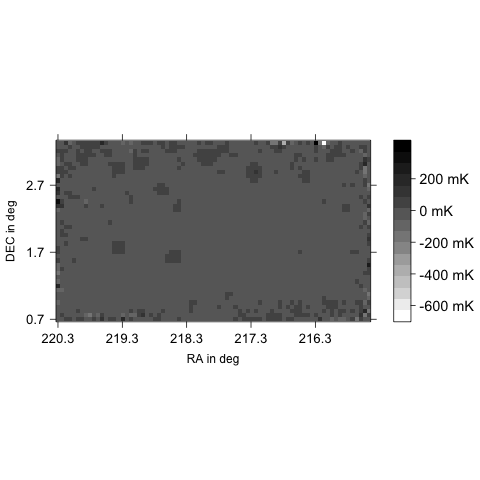}}\subfigure[6 ICs; masked]{
\includegraphics[width=0.5\textwidth, clip=true, trim= 0 100 0 130]
{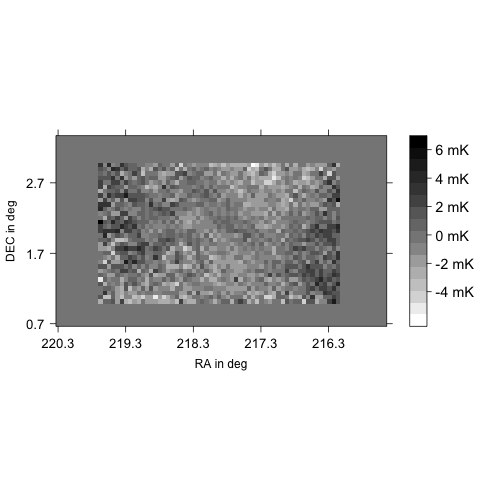}}
\subfigure[10 ICs]{
\includegraphics[width=0.5\textwidth, clip=true, trim= 0 100 0 130]
{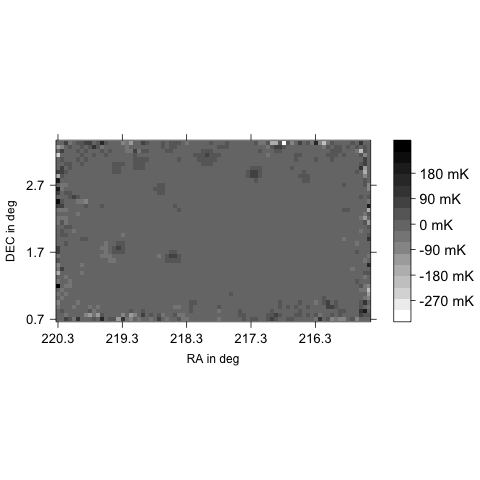}}\subfigure[10 ICs; masked]{
\includegraphics[width=0.5\textwidth, clip=true, trim= 0 100 0 130]
{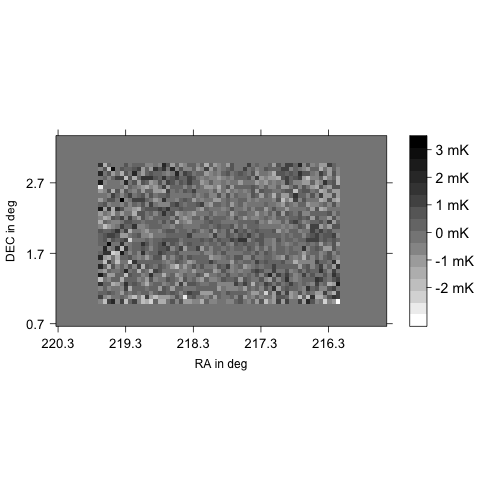}}
\caption{The sum of the temperature residual maps of sub-dataset A of
  the 15hr-field over all frequency channels.  The residual maps after
  the foreground removal with \textsc{fastica} should only contain
  noise and 21cm signal.  The first and second columns show the sum
  over the full field, and the analysis in which the edges of the
  fields are masked, respectively.}
 \label{summaps}
\end{figure*}

Furthermore, we examine the residual maps for point source
contamination at all frequencies.  This can be checked most accurately
by summing the residual maps over all frequencies. The
{instrumental noise and cosmological signal are
  expected to be close to Gaussian-distributed hence to show no
spatial structure when averaging over many frequencies, see e.g. \cite{1998ApJ...500L..83T, Baccigalupi:2000xy}}.  In
Fig.~\ref{summaps}, the frequency-combined residual map of the
sub-dataset A of the 15hr-field is shown for different numbers of ICs.
The analysis of the full field is shown in the first column, and the
masked field in the second column.  The results from the full maps
demonstrate again how the high noise at the edges of the field is not
fully modelled by \textsc{fastica}.  For the masked analysis with two
ICs, as seen in panel (b), the frequency-combined maps contain
point-source residuals.  These residuals fade out with increasing
number of ICs, until they are clearly removed for 10 ICs in panel (f).
These tests evidence how \textsc{fastica} is able to model and
subtract the strong point sources from the intensity maps using
$N_{\rm IC} \ge 6$.

The 1hr-field contains fewer strong point sources, but the
observations suffer from inhomogeneous noise properties due to shorter
integration times.  Although \textsc{fastica} can not effectively
model systematic effects with near-Gaussian distributions, the
individual sub-datasets exhibit different noise imprints such that
their effect on the cross-correlation between the maps should be
diminished.

\subsection{Calibration or instrumental resonance}

\begin{figure*}
\subfigure[res.; 2 ICs ]{
 \includegraphics[width=0.3\textwidth]
{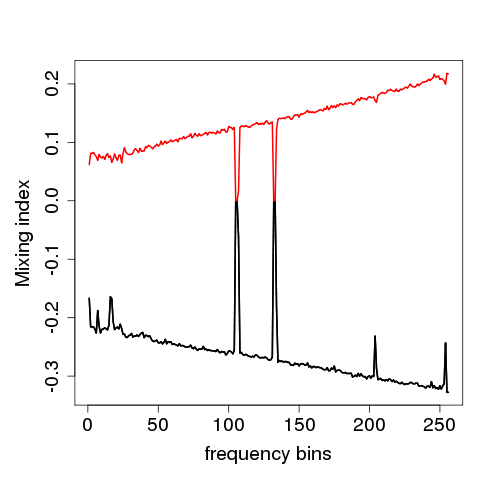}
}\subfigure[res.; 6 ICs ]{
 \includegraphics[width=0.3\textwidth]
{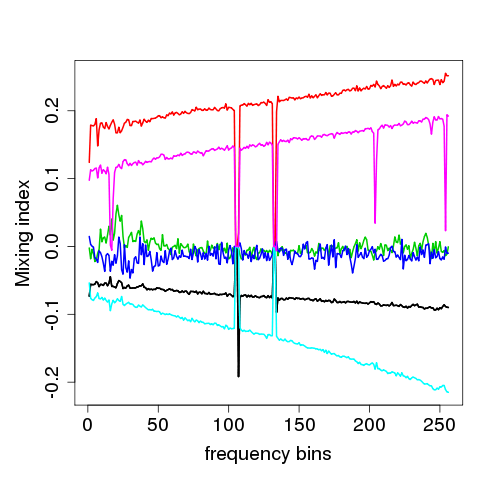}
}\subfigure[res.; 10 ICs]{
 \includegraphics[width=0.3\textwidth]
{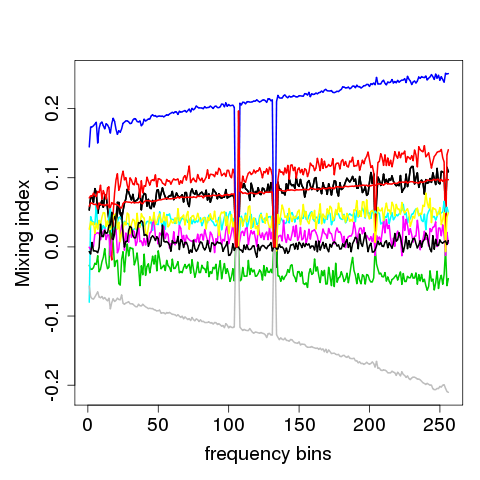}
}

\subfigure[no res.; 2 ICs]{
 \includegraphics[width=0.3\textwidth]
{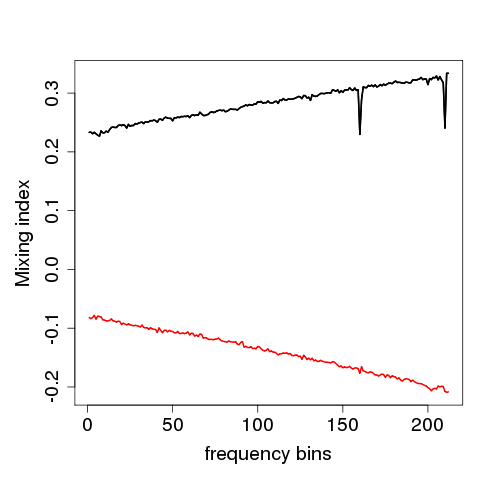}
}\subfigure[no res.; 6 ICs]{
 \includegraphics[width=0.3\textwidth]
{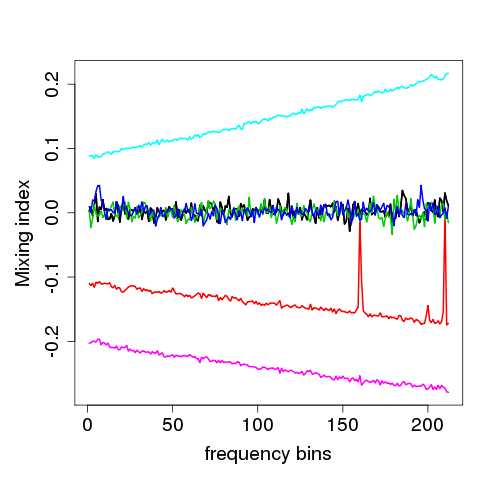}
}\subfigure[no res.; 10 ICs]{
 \includegraphics[width=0.3\textwidth]
{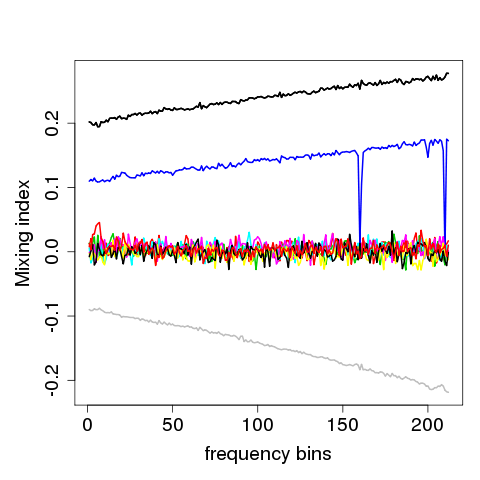}
}
\subfigure[no res.; masked; 2 ICs]
{\includegraphics[width=0.3\textwidth]
{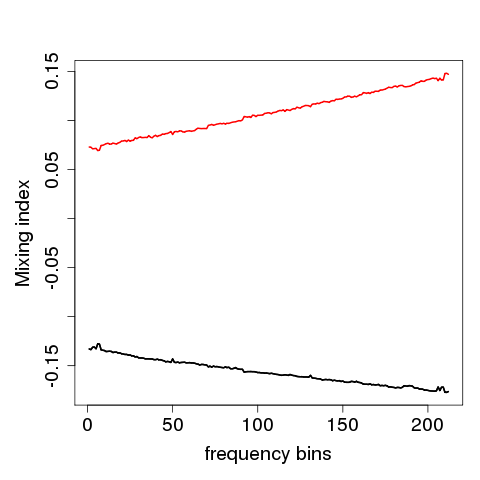}
}\subfigure[no res.; masked; 6 ICs]
{\includegraphics[width=0.3\textwidth]
{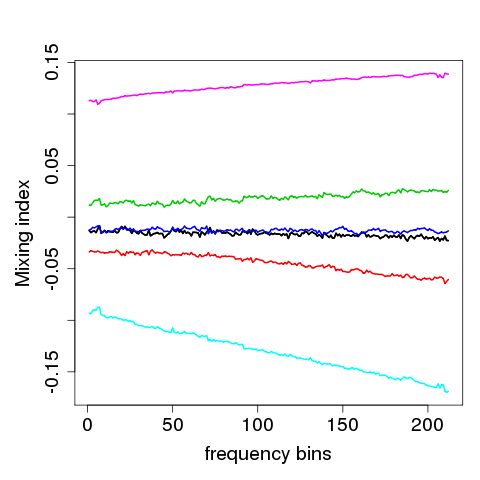}
}\subfigure[no res.; masked; 10 ICs]
{\includegraphics[width=0.3\textwidth]
{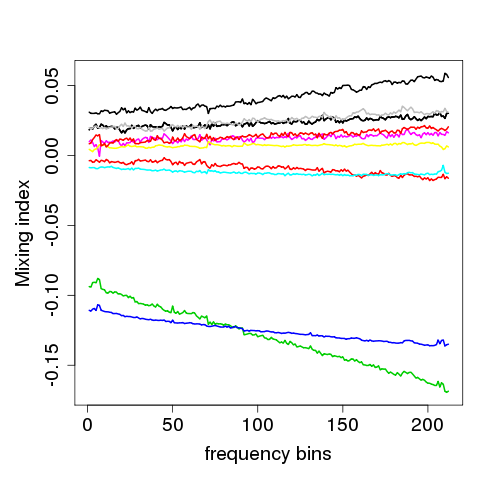}
}
\caption{The mixing matrix $\mathbf A$ as a function of frequency for
  an analysis of sub-dataset A of the 15hr-field.  The three columns
  show analyses using 2, 6 and 10 ICs, respectively.  The first row
  results from an analysis of the full field.  For the analysis shown
  in the second row, the frequency channels contaminated by instrument
  resonance, and the first few frequency channels, have been removed.
  In the third row, the-noise dominated edges of each map are also
  masked out, which produces a smoothly-varying variation of the
  mixing matrix with frequency, as expected in a successful foreground
  subtraction.}
\label{mixmode}
\end{figure*}

Some frequency channels of the GBT telescope are sensitive to
telescope resonance or RFI.  In addition, the calibration of the
telescope is a source of error in the amplitude of the measurements.
In Fig.~\ref{mixmode} the mixing modes $\boldsymbol{a_i}$, which give
the mixing amplitude per frequency channel, are plotted for an
analysis of one dataset of the 15hr-field with 2, 6 and 10 ICs as a
function of frequency bin, where bin 0 refers to $f = 900\rm MHz$,
i.e. $z=0.58$.  In a perfect foreground subtraction scenario, each
line should represent the flat spectral index of a foreground
component.  However, instrumental effects such as calibration errors,
varying thermal noise, frequency-dependent polarization errors and
telescope resonances disturb the flat spectra and allow identification
of corrupted data.  Around the frequencies 798 MHz and 817 MHz, two
known telescope resonances corrupt the measurements and are flagged
during the map-making.  These channels can be seen as the spikes in
panels (a), (b) and (c) of Fig.~\ref{mixmode}, where we performed
\textsc{fastica} on the full data set.  After removal of both the
contaminated frequency channels and the first few frequency bins which
show anomalies due to calibration uncertainties, the resulting mixing
modes are shown in panels (d), (e) and (f).  One mixing mode spectrum
for 2 ICs exhibits two features at high frequency bins which points to
a further irregularity in the data due to instrumental effects.  In
panels (e) and (f), using 6 and 10 ICs, we observe that some modes
show high fluctuations around a flat spectrum.  These large-amplitude
oscillations are due to \textsc{fastica} modelling dominant noise
features as ICs.  Again, excluding the noisy edges of the field solves
this issue, producing the results shown in panels (g), (h) and (i), in
which the mixing modes are relatively flat and featureless.  We are
therefore confident that \textsc{fastica} predominately identifies
frequency-dependent foreground components in this case, and we utilize
the masked 15hr field with $58 \times 33$ pixels in the remainder of
this work.

The analysis of the 1hr field exhibits similar improvements when
masking the edges, although more fluctuations in the mixing modes are
obtained as \textsc{fastica} attempts to model the strong noise
features present in these observations.  Our default mask of the
1hr-field results in $121 \times 53$ pixels.

\subsection{Noise properties }

\begin{figure}
\subfigure[2 ICs]{
\includegraphics[width=0.5\textwidth, clip=true, trim= 0 100 0 130]
{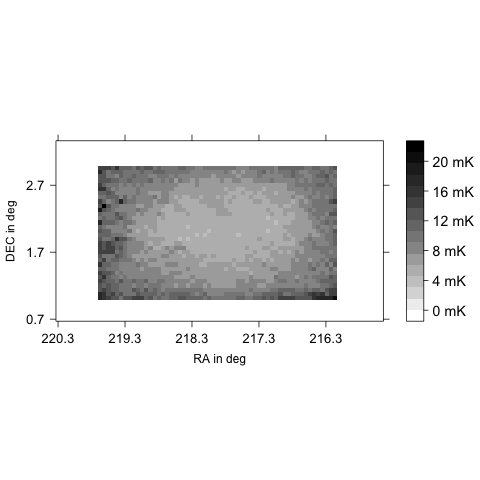}}
\subfigure[6 ICs]{
\includegraphics[width=0.5\textwidth, clip=true, trim= 0 100 0 130]
{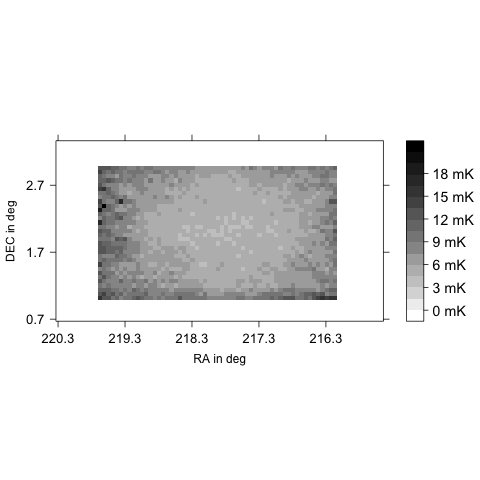}}
\subfigure[10 ICs]{
\includegraphics[width=0.5\textwidth, clip=true, trim= 0 100 0 130]
{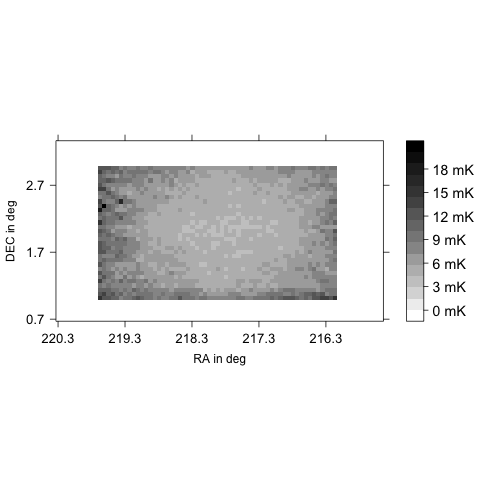}}
\caption{The standard deviation of the temperature residual maps
  across the frequency channels for sub-dataset A of the 15hr-field,
  for analyses using 2, 6 and 10 ICs.  The standard deviation is
  unaffected by the number of ICs chosen.}
 \label{sdmaps}
\end{figure}

In Fig.~\ref{sdmaps} we show the standard deviation of the residual
maps along the line of sight.  For noise-dominated data we expect the
standard deviation to be much higher than the amplitude of the sum of
all pixels, as can be seen when comparing the standard deviation
values with Fig.~\ref{summaps}.  The structure of the standard
deviation maps additionally shows how the noise varies with spatial
position.  For sub-dataset A of the 15hr-field in Fig.~\ref{sdmaps},
this structure is stable when increasing $N_{\rm IC}$ from 2 to 6 and
10.  This suggests that the leakage of noise into the reconstructed
Galactic foregrounds is low, and confirms the Gaussian nature of the
instrumental noise.

The noise levels of the residuals of the 1hr-field maps are less
stable than the 15hr-field with increasing number of ICs, and are
dominated by single features with irregular distribution over the map
due to the differing observational depth of the
pixels. \textsc{fastica} can incorporate some of the strong features
as ICs, partially removing the noise systematics.  We note that the
noise structure of each sub-dataset differs, which prevents
contamination of the cross-correlation.

We can access more information about the structure of the data by
measuring the 2D power spectra of maps corresponding to individual
frequency channels.  Following the formalism of Sec.~\ref{PS},
Fourier-transformed temperature maps are calculated as $\tilde
T_A(\vec k_l) = \sum_{j=1}^{N_{pix}} T_A(\vec x_j) \exp{(i \vec k_l
  \cdot \vec x_j)}$, and the 2D power spectrum is defined as
$P_{2D}(\vec k_l) = \mathrm{Re}\{\tilde T_A(\vec k_l)\tilde T_B(\vec
k_l)^*\}$.  The noise power spectrum of a frequency map can be
estimated via two measures:
\begin{itemize}
 \item A jack-knife test: the difference of two sub-datasets should
   only contain thermal noise, since the astrophysical signal remains
   unchanged with time.  We obtain an estimate of the noise power
   spectrum of one map by calculating the power spectrum of the
   difference map and dividing it by two.  The difference maps also
   encode systematic errors between the sub-datasets.
\item Auto-correlation: the power spectrum of each sub-dataset after
  the foreground removal should be a proxy for the noise, if it
  dominates the HI signal.
\end{itemize}

\begin{figure*}
\centering
\subfigure[$\nu=838\rm MHz$]{ \includegraphics[width=0.3\textwidth, clip=true, trim= 260 140 280 150]
{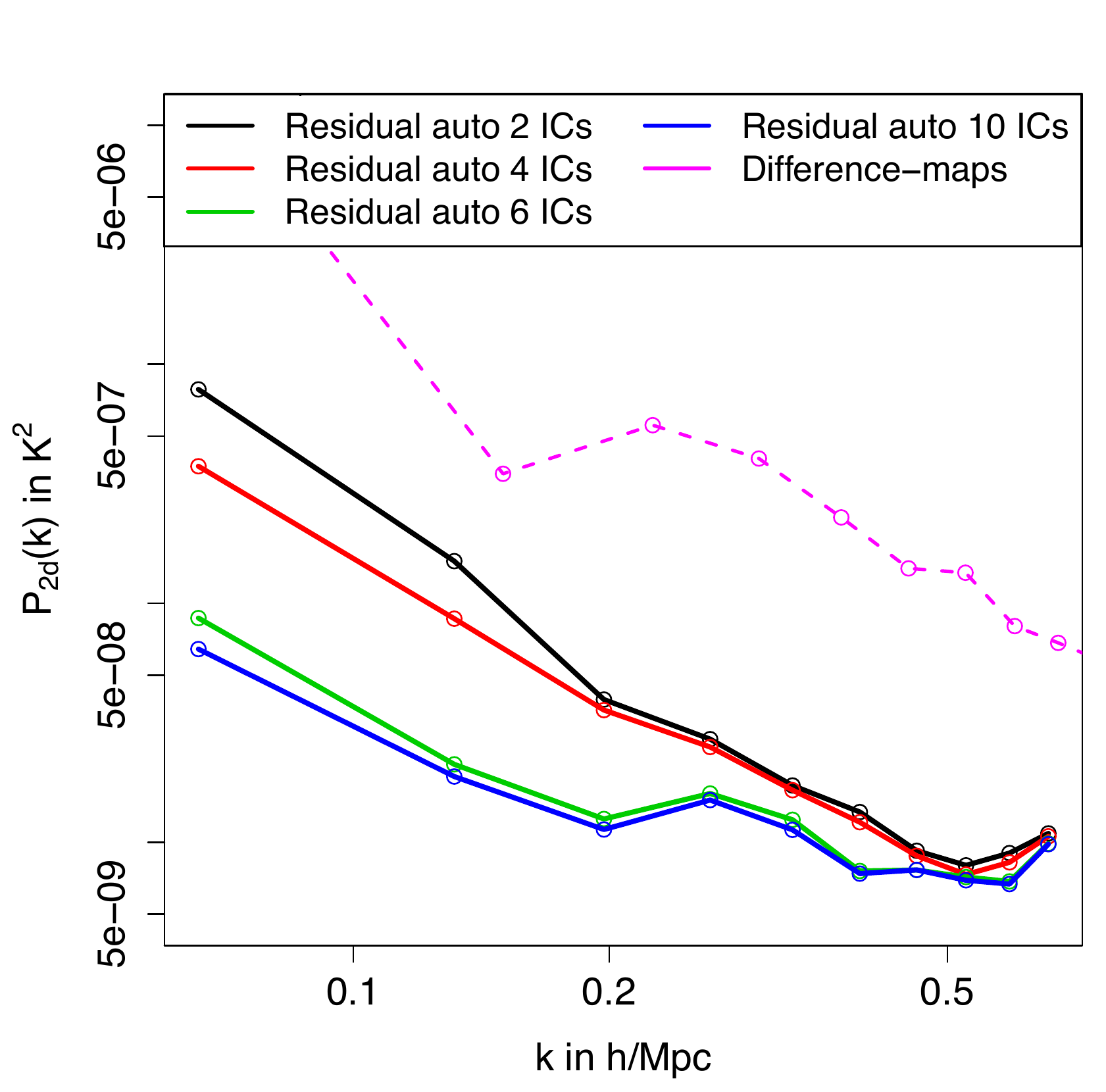}}
\subfigure[$\nu=806\rm MHz$]{  \includegraphics[width=0.3\textwidth, clip=true, trim=260 140 280 150]
{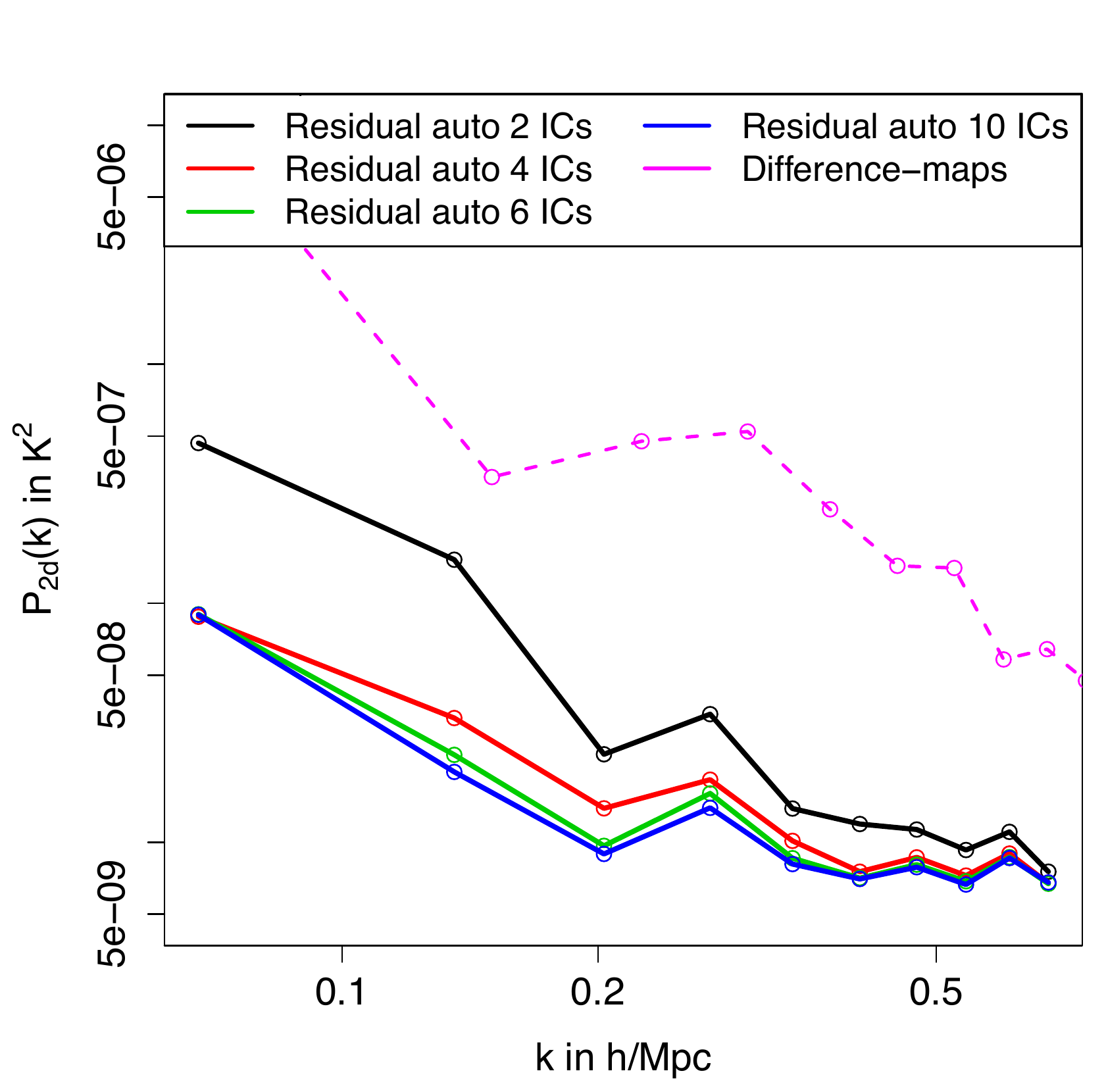}}
\subfigure[$\nu=775\rm MHz$]{  \includegraphics[width=0.3\textwidth, clip=true, trim=260 140 280 150]
{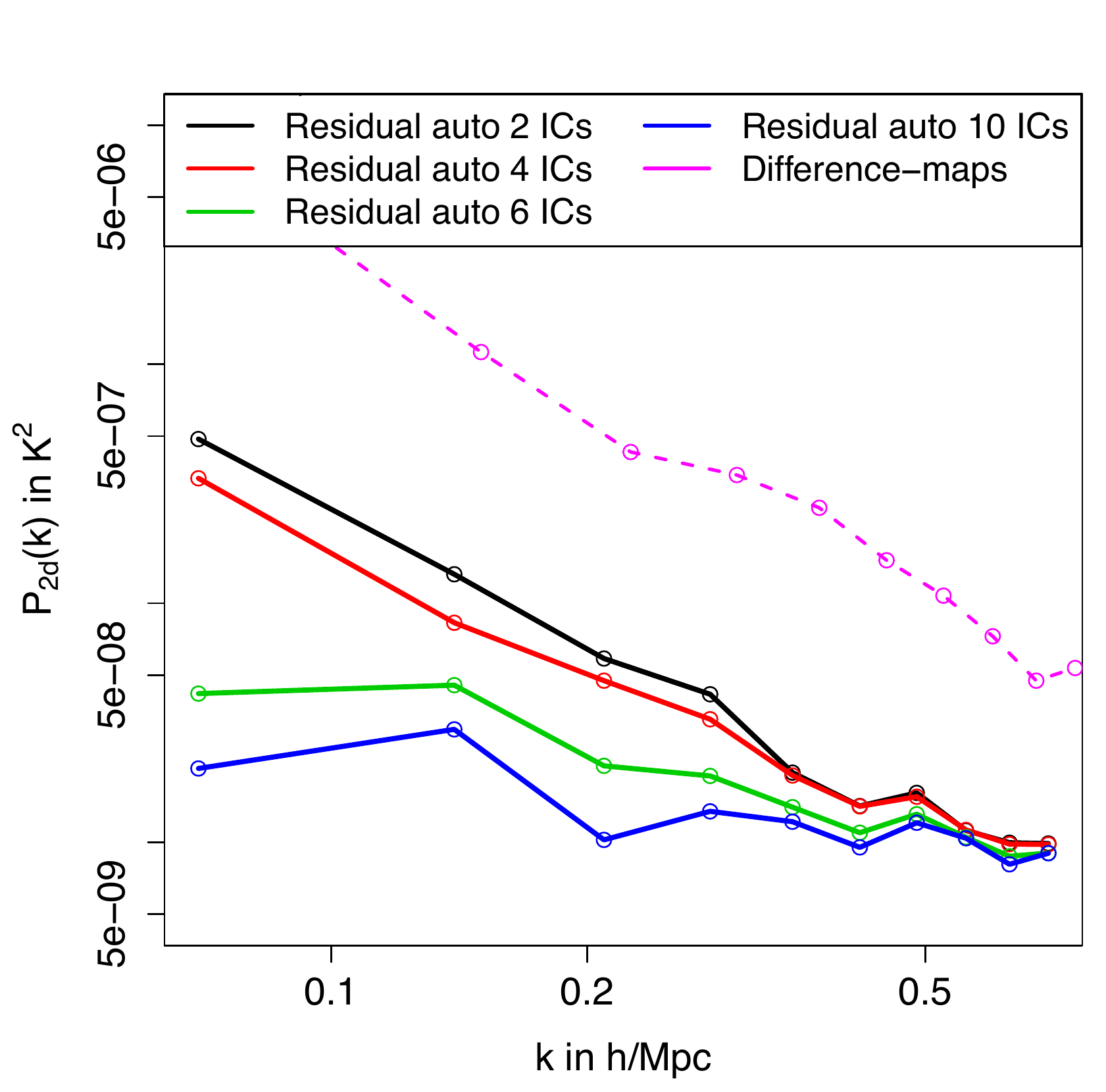}}
\caption{2D auto-power spectra of the residual maps of the 15hr-field
  for different ICs (solid lines), in comparison to the difference-map
  power spectrum of the original maps (dashed lines).  Each power
  spectrum is an average of all possible combinations of sub-datasets.
  These power spectra are a noise approximation of the maps before and
  after the foreground removal.}
\label{noisecorr}
\end{figure*}

\begin{figure*}
\subfigure[Sub-dataset (A - B)]{
\includegraphics[width=0.33\textwidth, clip=true, trim= 0 110 0 110]
{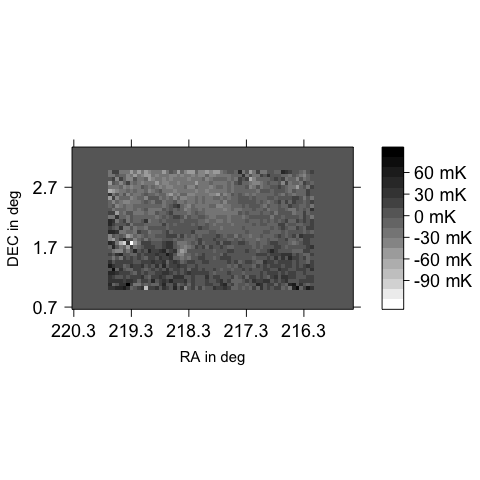}
}\subfigure[Sub-dataset (A - C)]{\includegraphics[width=0.33\textwidth, clip=true,
trim= 0 110 0 110]
{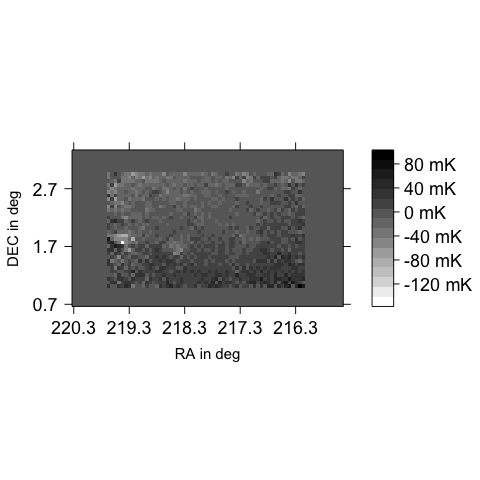}}\subfigure[Sub-dataset (B - C)]{\includegraphics[
width=0.33\textwidth , clip=true, trim= 0 110 0
110]{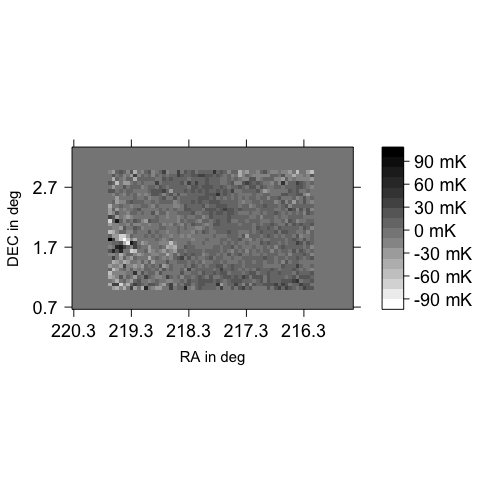}}
\caption[The difference maps of the 15hr-field observations]{The
  difference maps of the 15hr-field observations for frequency $\nu =
  869 \rm MHz$. The difference maps exhibit spatial structure due to
  systematic errors in the sub-datasets.}
\label{diffmaps}
\end{figure*}

In Fig.~\ref{noisecorr} we show a few examples of the 2D power spectra
of the difference maps and the auto-correlations of the 15hr-field,
where we averaged over all possible combinations of sub-datasets.  It
can be seen that the difference-map correlations contain more power
than the auto-correlations. This can be explained in terms of the
spatial structure of the difference maps, shown in
Fig.~\ref{diffmaps}.  The difference maps show a clear structure at
the position of the point sources, produced by instrumental effects
which correlate with the amplitude of signal such as calibration
errors, pointing offsets and thermal noise.  The amplitude of this
systematic contribution does not depend on frequency.
\textsc{fastica} models the point sources in each sub-dataset
independently, hence can remove these systematic effects.  The
analysis of the 1hr-field shows similar behaviour.

\subsection{Residual-foreground correlation}

\begin{figure*}
\subfigure[Frequency bins 80-99;  no mask]{
\includegraphics[width=0.5\textwidth, clip=true, trim= 125 150 130 150]
{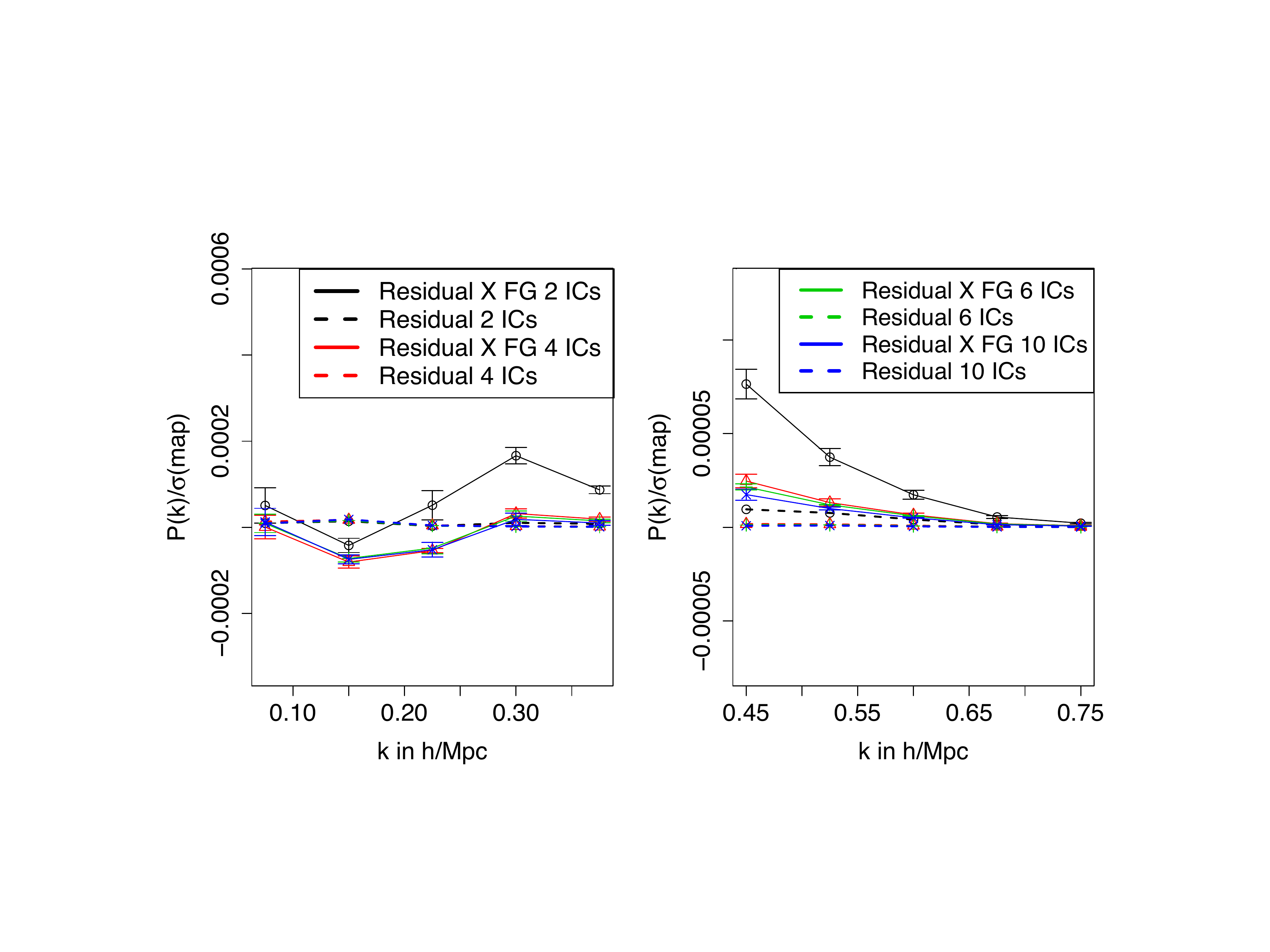}
}\subfigure[Frequency bins 80-99; edges masked]{
 \includegraphics[width=0.5\textwidth, clip=true, trim= 125 150 130 150]
{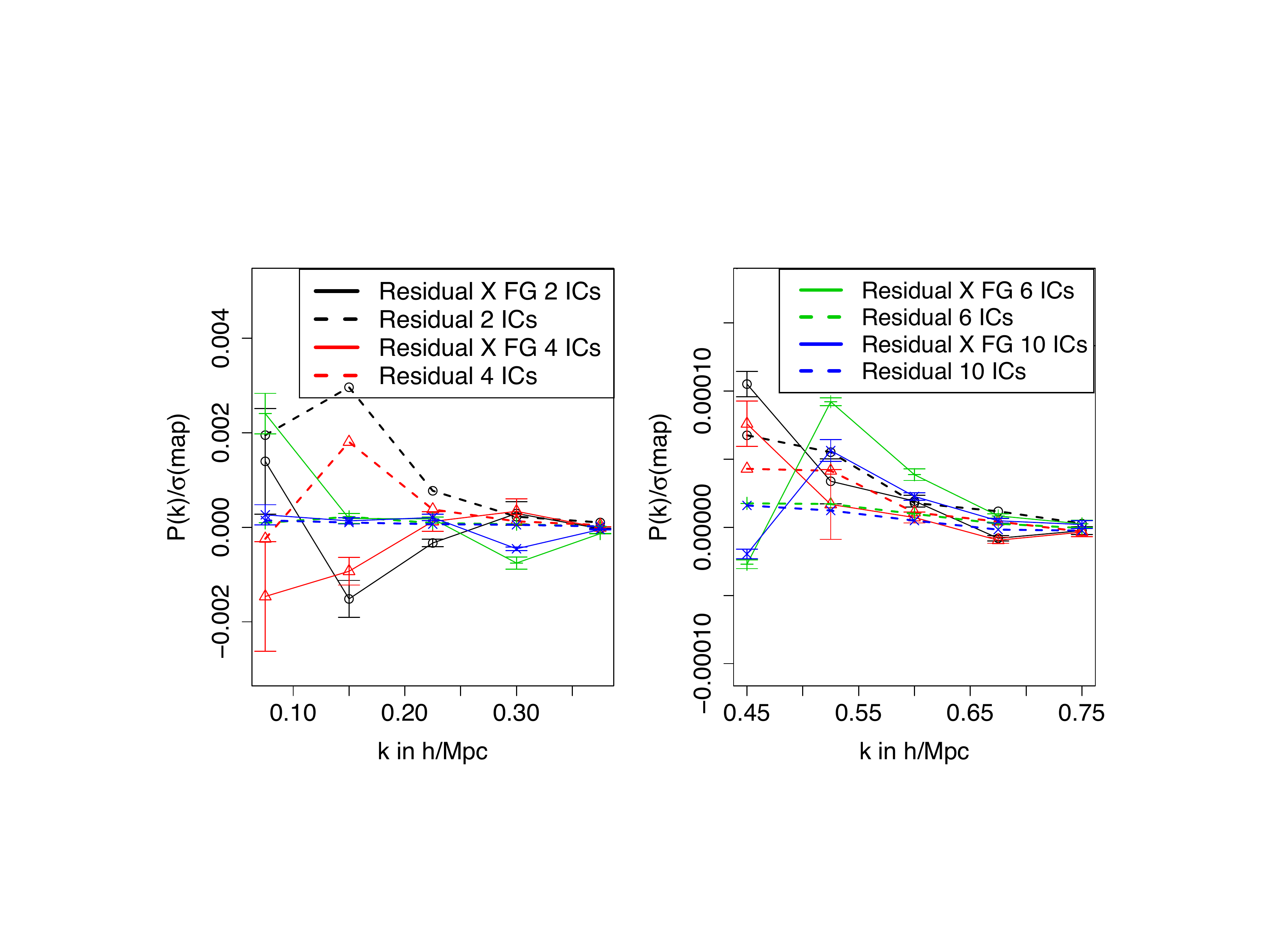}}
\subfigure[Frequency bins 140-159; no mask]{
\includegraphics[width=0.5\textwidth, clip=true, trim= 125 150 130 150]
{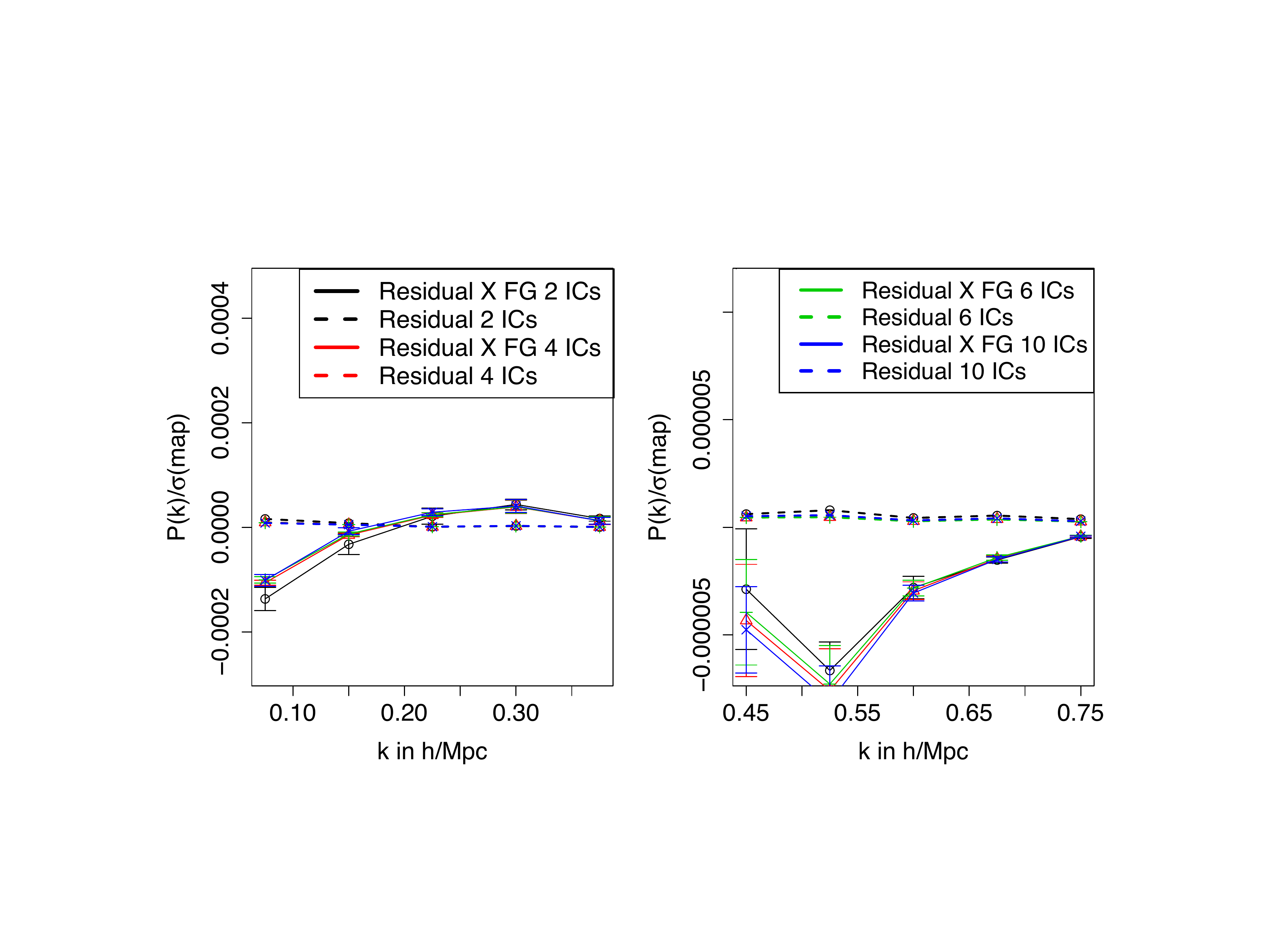}
}\subfigure[Frequency bins 140-159; edges masked]{
 \includegraphics[width=0.5\textwidth, clip=true, trim= 125 150 130 150]
{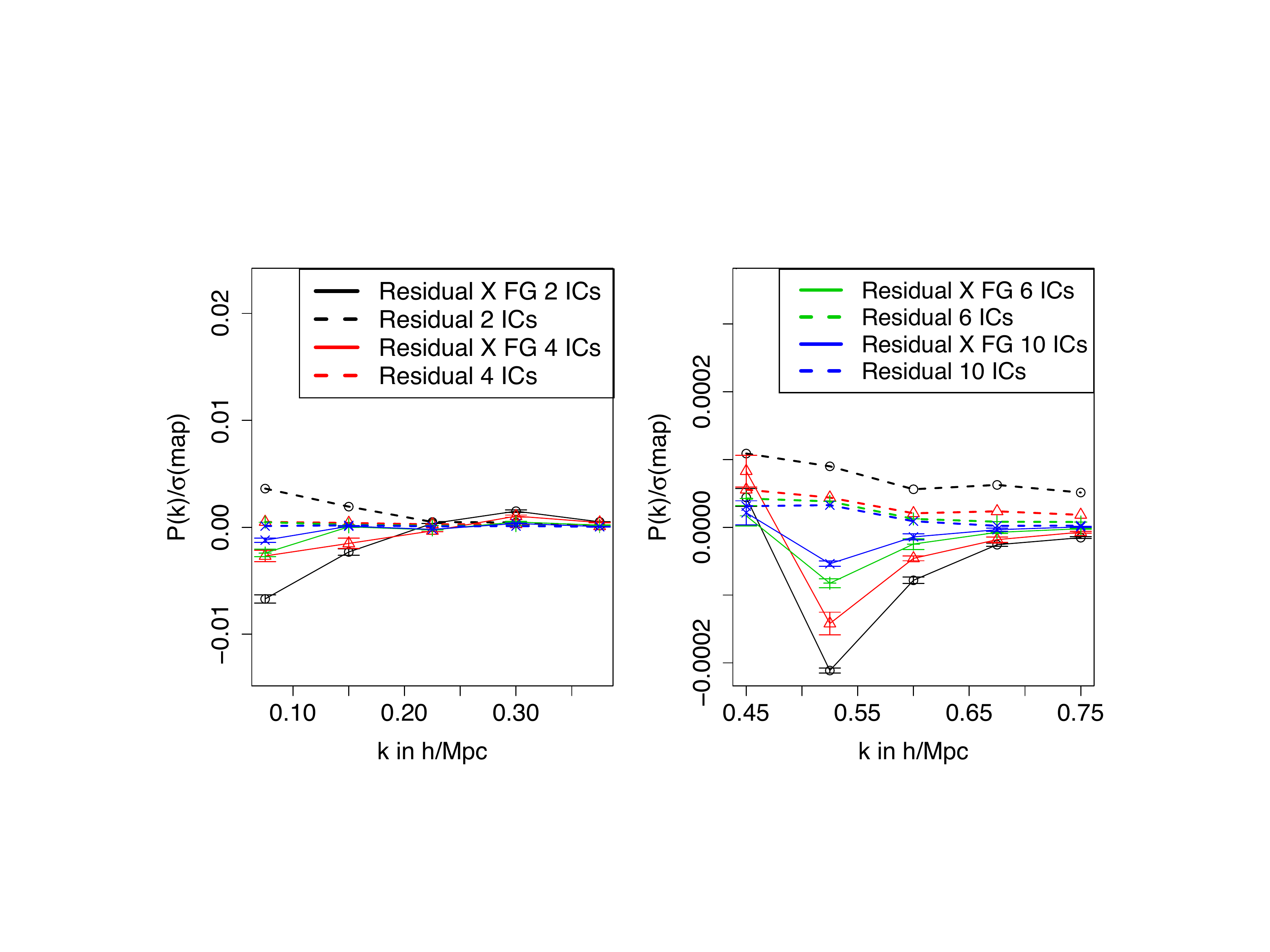}
}
\subfigure[Frequency bins 220-239; no mask]{
\includegraphics[width=0.5\textwidth, clip=true, trim= 125 150 130 150]
{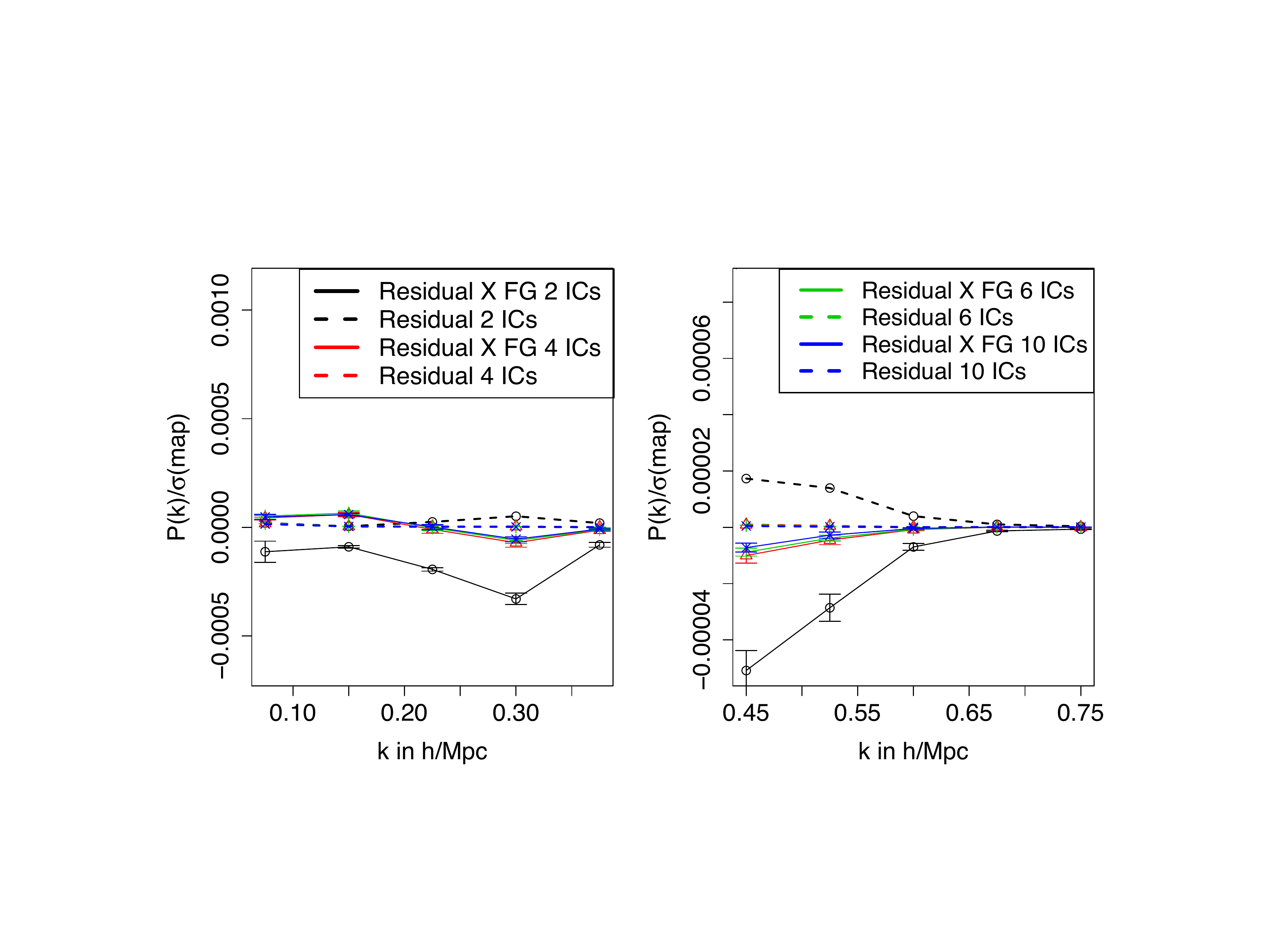}
}\subfigure[Frequency bins 220-239; edges masked]{
 \includegraphics[width=0.5\textwidth, clip=true, trim= 125 150 130 150]
{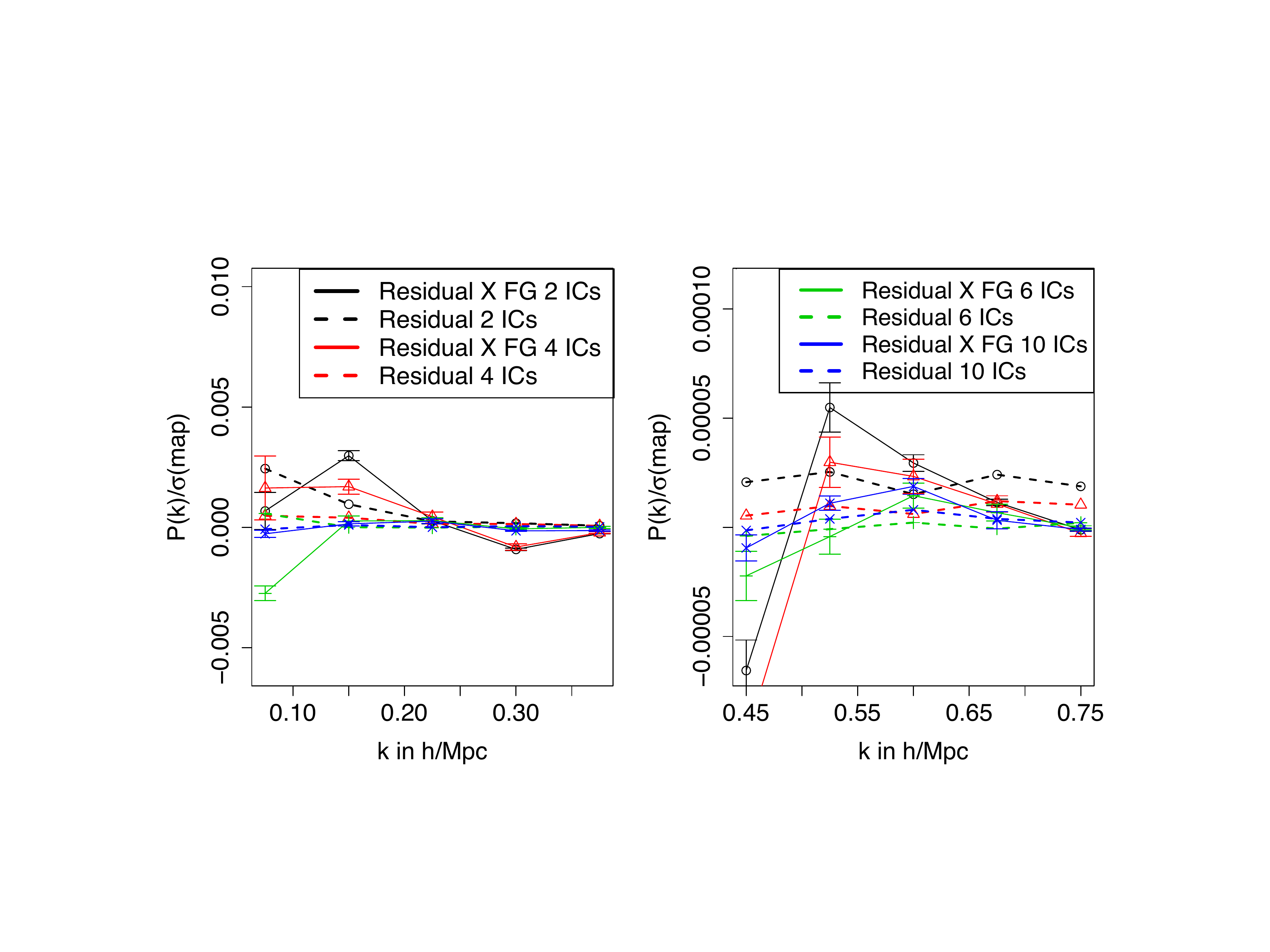}
}
\caption[2D cross-correlation between foregrounds and residuals of the
  15hr-field]{The 2D cross-power spectra between reconstructed
  foregrounds and residuals of the 15hr-field, marked as the solid
  lines for different numbers of ICs with different colors.  The
  dashed lines represent the residual power spectrum between
  sub-datasets.  Each panel shows the 2D power spectrum with different
  scales for large and small wavenumbers $k$.  The correlations in
  each row are estimated over three frequency ranges, each containing
  20 frequency bins.  The first column of 2 panels is for a full-field
  analysis, and the second column shows results for the masked field.}
\label{2dcrosscorr15hr}
\end{figure*}

We can also evaluate the foreground removal by considering the 2D
cross-power spectra between different frequency maps. In the following
plots, we show two kinds of correlations:
\begin{itemize}
\item The cross-correlation of the residual maps from different
  sub-datasets.  This cross-correlation should be driven by the
  cosmological signal, since the noise is uncorrelated between
  sub-datasets.  However, it could also be produced by residual
  foreground contamination.
\item The cross-correlation of the residuals and the reconstructed
  foreground maps.  Such a signal could be produced if the foregrounds
  are insufficiently modelled and contaminate the residuals, or if the
  ICs contain instrumental noise or cosmological signal.
\end{itemize}
In Fig.~\ref{2dcrosscorr15hr} we display examples of these 2D power
spectra, analyzing the full field in the first column and the masked
field in the second column, for a series of stacks of 20 frequency
channels.  Since the maps are dominated by thermal noise, the noise
decreases as $1/\sqrt{N}$ when adding $N$ frequency channels.

The amplitudes of the cross-correlations are proportional to the
product of mean temperatures of the two input maps.  In order to
compare the cross-correlation of foreground and residuals to the
correlation of residual maps we need to normalize the amplitude, for
which we use the standard deviations of the respective maps.

In the first column, the figures show the results of the flawed
\textsc{fastica} decomposition which insufficiently removes the point
sources of the foregrounds, as seen in the stacked maps in
Fig.~\ref{summaps}.  The solid lines show a similar behaviour for
different number of ICs and frequencies, indicating a correlation
between foregrounds and residuals.  In the second column the masked
results, which are clean of point source contamination, are shown. The
cross-correlation of foreground and residuals are relatively
random-distributed and are an indication that the data is dominated by
statistical noise not systematic foregrounds.  The dashed lines in all
figures are the residual correlation between all combinations of
sub-datasets.  These converge for increasing number of ICs, confirming
the results of the successful foreground removal of previous tests and
demonstrating that our results do not sensitively depend on the number
of ICs chosen.  The cross-correlation of the residuals and foregrounds
of the 1hr-field show similar behaviour.

\section{3D Power Spectrum Results}
\label{res}

\subsection{Auto-Correlations}

\begin{figure*}
\subfigure[15hr-field]{
\includegraphics[width=0.4\textwidth, clip=true, trim= 0 10 0 50]
{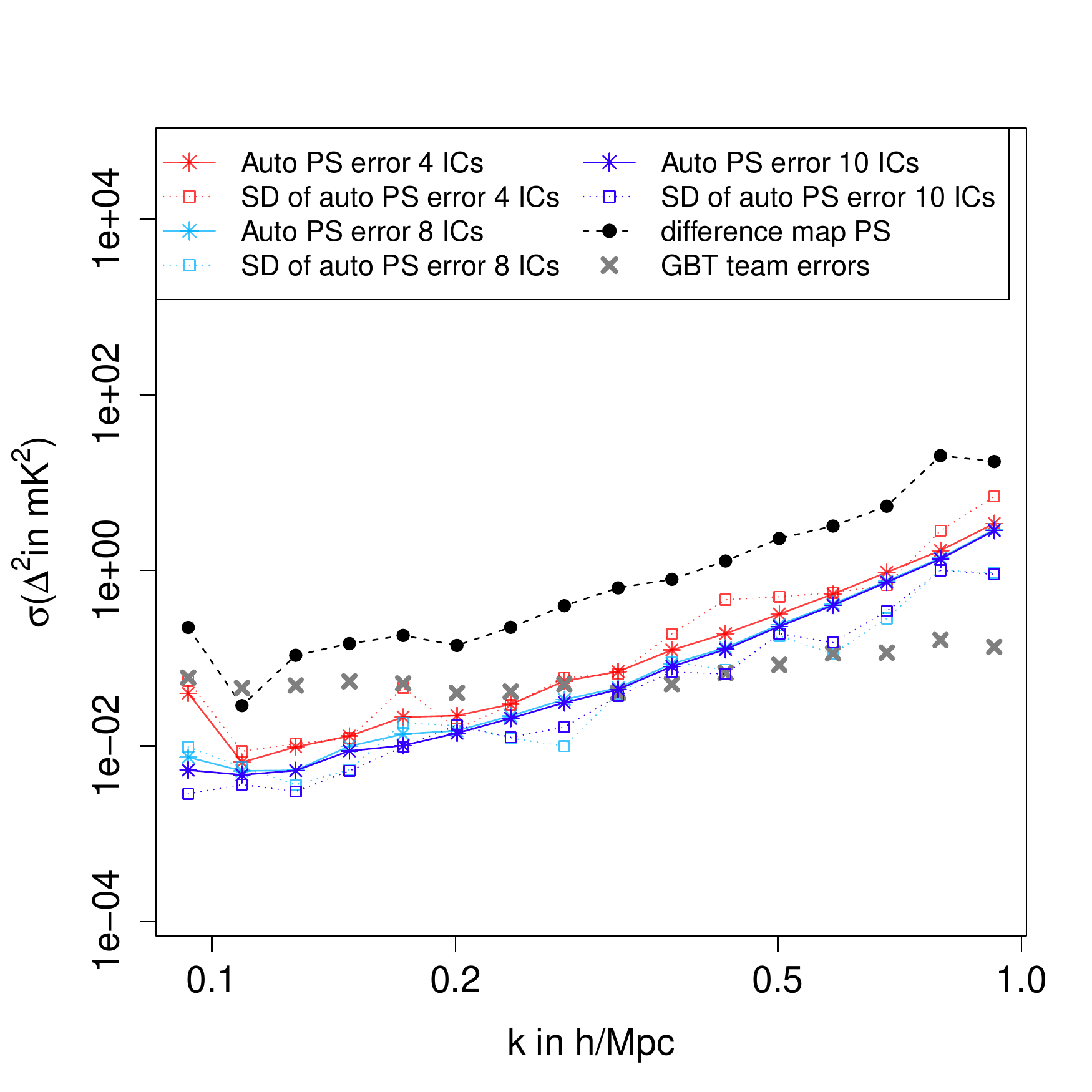}
}\subfigure[1hr-field]{
\includegraphics[width=0.4\textwidth, clip=true, trim= 0 10 0 50]
{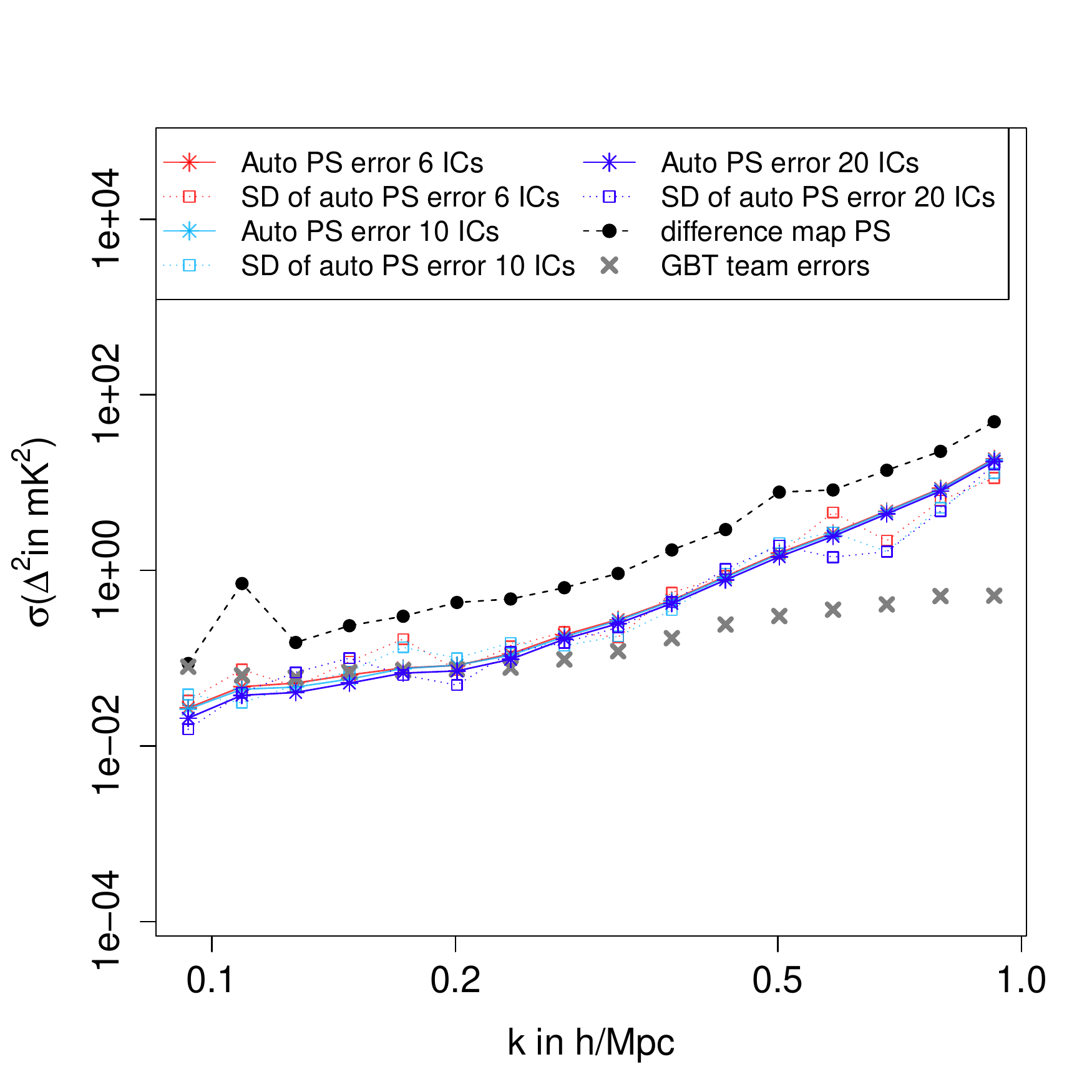}
}
\caption[Error estimates on 3D power spectrum measurement]{Different
  estimates of the error in the 3D power spectrum measurement are
  shown for different numbers of ICs, for the 15hr-field and
  1hr-field.  The black dashed line is the error estimate based on the
  difference maps.  The solid coloured line shows the error based on
  the auto-correlations of the sub-dataset, and the dotted coloured
  line represents the standard deviation of the cross-power spectrum
  measurements between the sub-datasets.  The error estimates of the
  SW13 analysis are marked by grey crosses.}
\label{3dPSGBTerror}
\end{figure*}

In this section we present the results of the 3D power spectrum
estimation from the GBT intensity maps.  We consider three different
strategies for estimating errors in the power spectrum measurements,
and compare these in Fig.~\ref{3dPSGBTerror}:
\begin{itemize}
 \item We use the auto-correlations of the residual maps as a proxy
   for the noise power spectrum in Eq.~\ref{autocorrnoise} (solid
   error bar).
\item We use the power spectrum of the difference of the maps, divided
  by 2, as a proxy for the noise power spectrum in
  Eq.~\ref{autocorrnoise} (dashed error bar).
\item We calculate the standard deviation of the 6 sub-dataset
  cross-power spectra used in the analysis, and divide it by $\sqrt{6}$
  to produce an error in the mean (dotted error bar).
\end{itemize}

The error based on the noise estimate from the difference maps (black
dashed line) is higher than the other two error estimates.  We believe
that this provides an upper limit on the error in the measurements
since it includes systematic effects correlated with the foregrounds,
which \textsc{fastica} partially subtracts from the data.  The noise
estimate from the auto-correlation gives a better approximation to the
errors in the foreground-subtracted measurements.

\begin{figure*}
\subfigure[15hr-field]{
\includegraphics[width=0.4\textwidth, clip=true, trim= 0 10 0 50]
{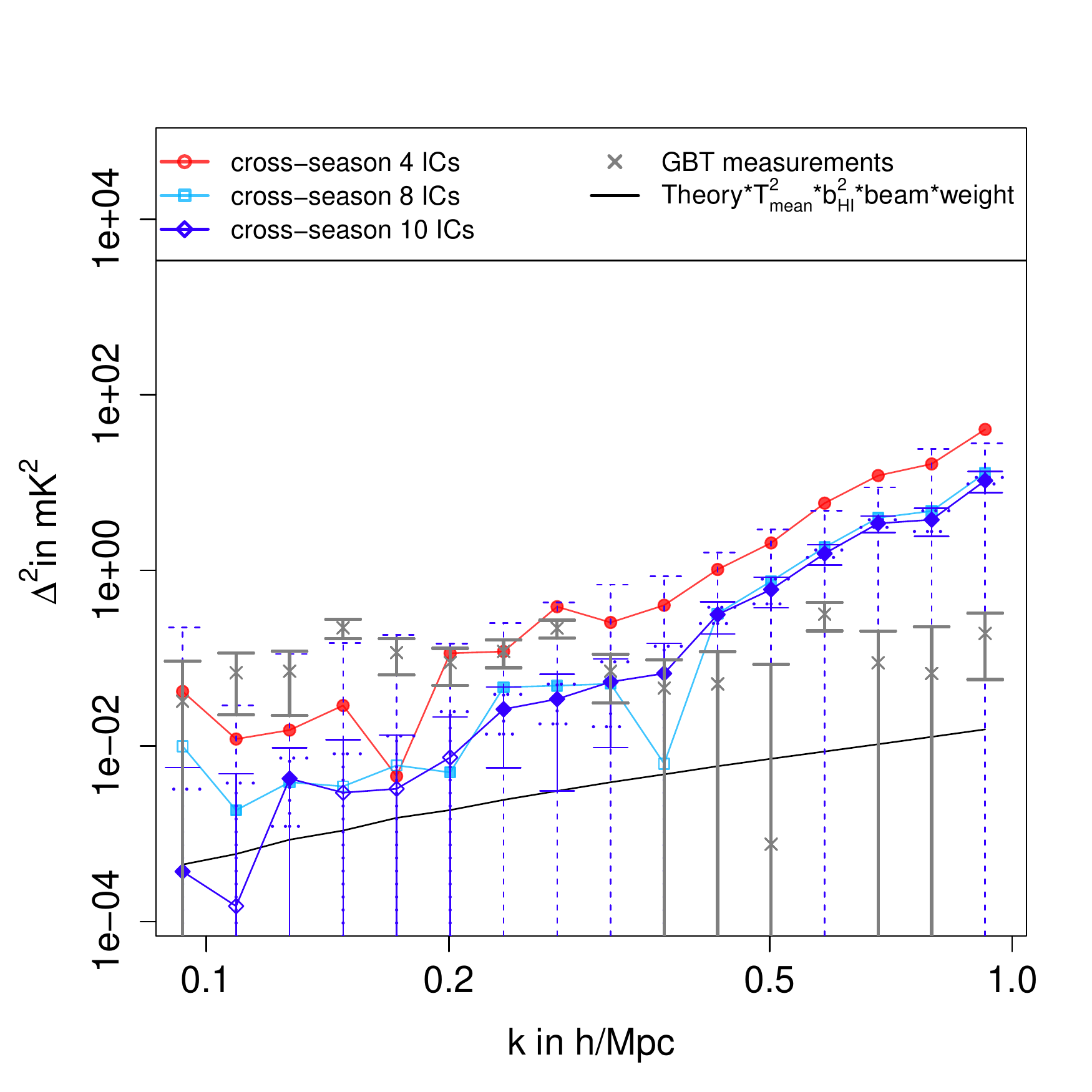}
\label{3dPSGBTa}
}\subfigure[1hr-field]{
\includegraphics[width=0.4\textwidth, clip=true, trim= 0 10 0 50]
{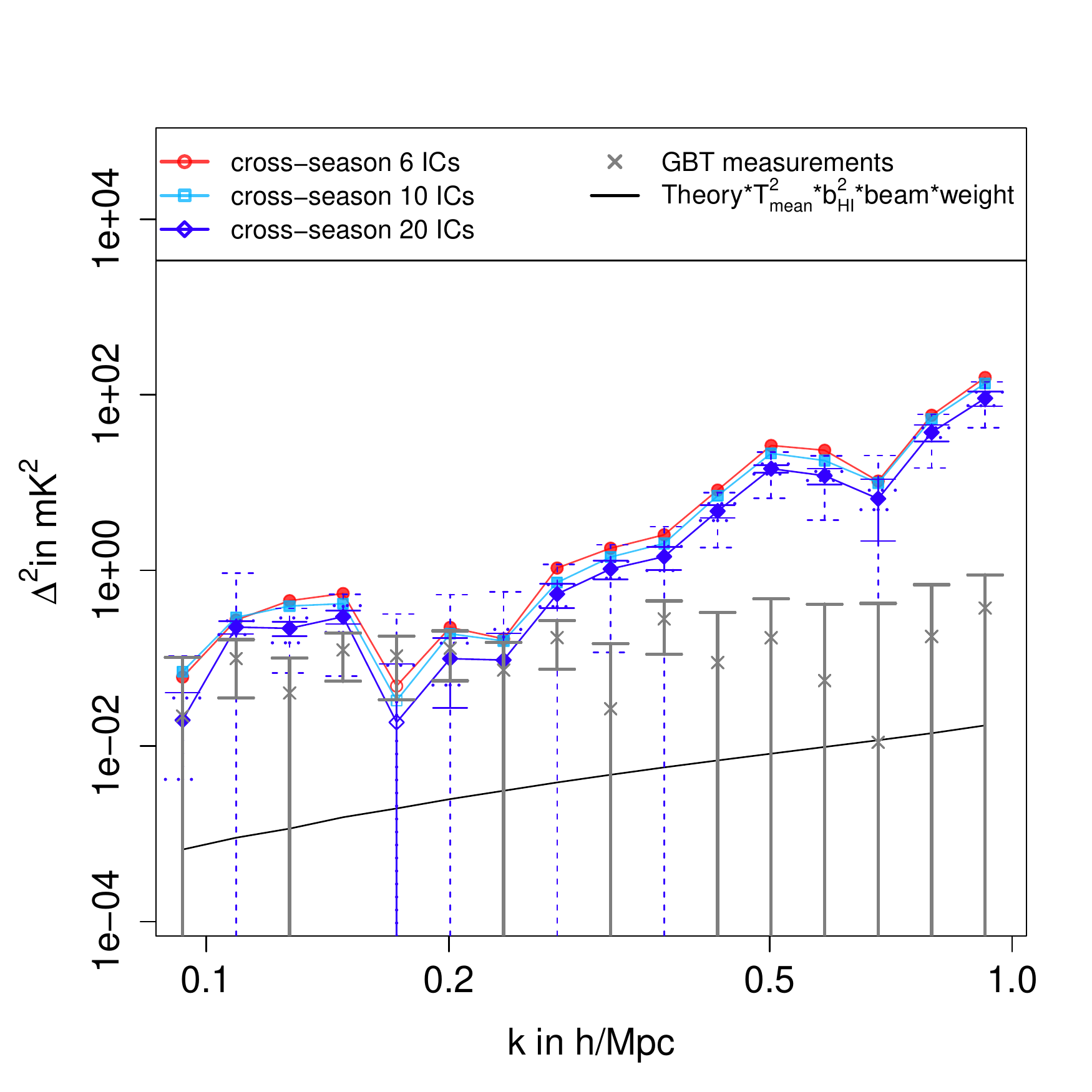}
\label{3dPSGBTb}
}
\caption{The 3D intensity mapping cross-power spectrum between
  sub-datasets of the 15hr field data (left-hand panel) and 1hr field
  (right-hand panel), showing the different error estimates from
  Fig.~\ref{3dPSGBTerror} using the respective line styles.  The
  measurements from SW13 are marked by grey crosses.  The black line
  shows the theoretical model power spectrum convolved with the window
  functions assuming $\Omega_{\rm HI}b_{\rm HI} = 0.43\cdot 10^{-3}$.}
\label{3dPSGBT}
\end{figure*}

In Fig.~\ref{3dPSGBT}, we present the intensity mapping power spectrum
estimates for different numbers of ICs used for the foreground
subtraction, in comparison with the results published by SW13, which
are marked by grey symbols.  The estimates are the average of all
possible combinations of the cross-power spectra between the 4
sub-datasets, showing the different error estimates from
Fig.~\ref{3dPSGBTerror} with their respective line
styles. {The power spectra have all been corrected for
  the telescope beam, using a constant beam model with $\theta_{\rm
    FWHM} = 0.44\deg$ for the SW13 data points, and a
  frequency-dependent beam for the \textsc{fastica} measurements.}

The power spectra converge with increasing number of ICs, showing that
\textsc{fastica} is a robust method to remove the non-Gaussian
foregrounds.  In Fig.~\ref{3dPSGBTa}, we see that our measured power
spectra in {both fields are in reasonable agreement
  with the results of SW13 on large scales with $k<0.2
  h\rm{Mpc}^{-1}$, but diverge for smaller scales.  The power spectrum
  amplitude of the 1hr-field is higher than the 15hr field, due to
  some residual foregrounds and significant instrumental systematics
  in the 1hr-field maps.}  The GBT measurements are corrected for
signal loss by an anisotropic transfer function $T(k_\perp,
k_\parallel)$, as described by \cite{Switzer:2015ria}.  The power
spectrum of the \textsc{fastica}-cleaned data does not require any
corrections by a transfer function since the signal loss is
negligible, as shown by \cite{Wolz:2013wna}.

The high amplitude of the intensity mapping power spectra measured by
\textsc{fastica} on smaller scales is driven by its conservative
approach to foreground subtraction.  This is in contrast to the SVD
method, which removes modes with high amplitudes regardless of their
statistical properties.  {This comparison shows that
  \textsc{fastica} provides a robust upper limit on the foreground removal, while
  SVD could provide a lower limit on the removable foreground modes.  Both methods have been
  shown to perform well in a simulated environment
  \citep{2015MNRAS.447..400A}.  However, in the presence of high
  instrumental noise and systematics, the foreground removal
  methodology can lead to significant differences.  \textsc{fastica}
  succeeds in removing resolved point sources and diffuse
  frequency-dependent foregrounds dominating on large scales.  However,
  it is not equipped to mitigate systematics on smaller scales
  dominated by thermal noise.  The SVD approach removes modes on all
  scales, but is prone to HI signal loss.  We believe that the
  application of both methods is a useful approach when investigating
  foregrounds and systematics of intensity mapping data.}

In general, the auto-correlations of the 15hr and 1hr fields are high
compared to the theoretical prediction.  This discrepancy could be
explained in several ways.  Systematics leftover from the foreground
subtraction could boost the amplitude of the power spectrum, and
additional power could be added to the 21cm signal by fluctuations
introduced by polarization leakage.  Finally, a different predicted
amplitude could be produced by changing the value of $\Omega_{\rm HI}
b_{\rm HI}$.

\subsection{Cross-Correlation with WiggleZ}

\begin{figure*}
\subfigure[15hr-field]{
\includegraphics[width=0.45\textwidth]
{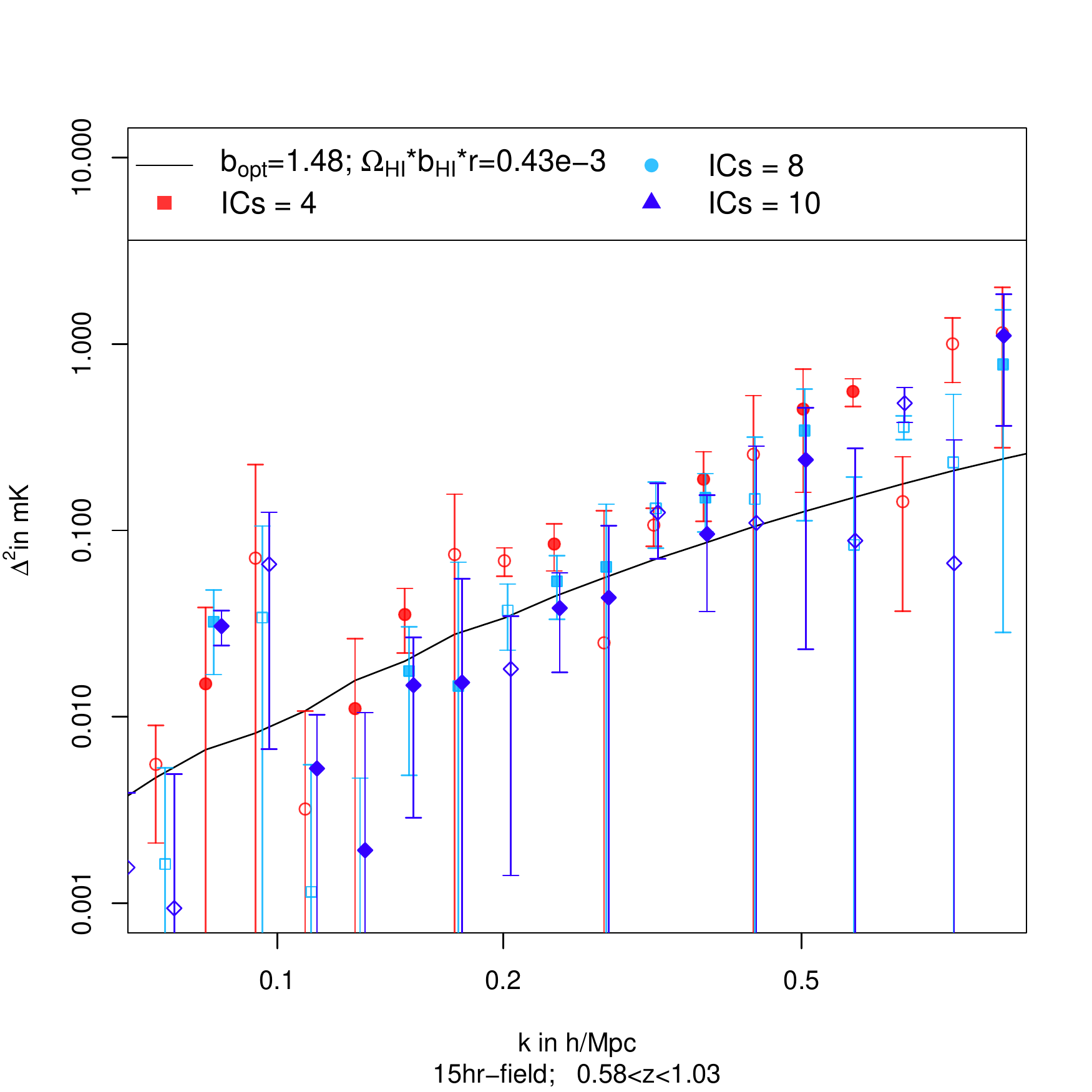}
}\subfigure[1hr-field]{
\includegraphics[width=0.45\textwidth]
{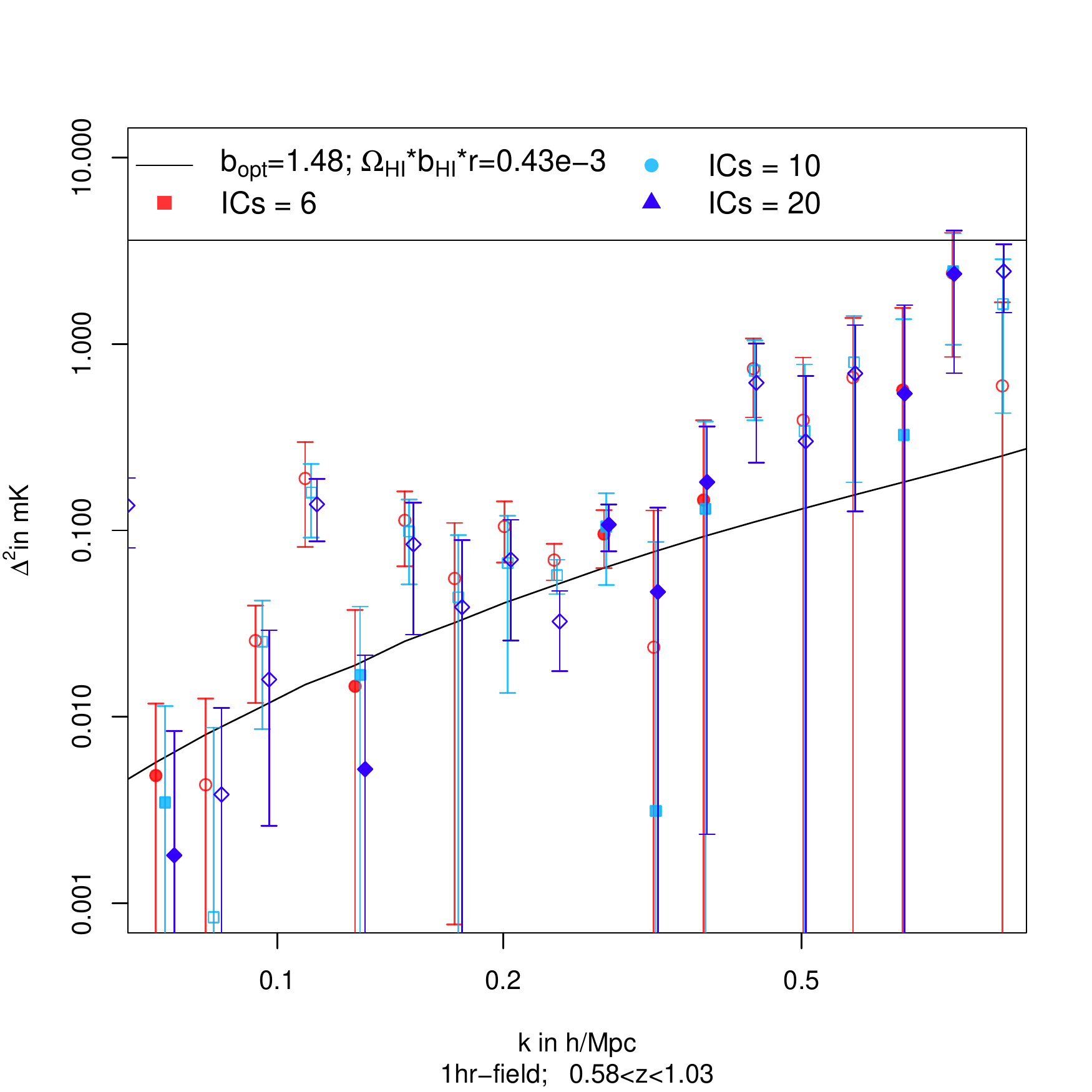}
}
\label{3dPSGBTXWIGICs}
\caption[]{The 3D cross-power spectrum of the GBT intensity maps and
  WiggleZ galaxies, for foreground subtraction with different numbers
  of ICs.  Results are shown for the 15hr-field in the left panel and
  the 1hr-field in the right panel.  The black lines are the
  theoretical model convolved with the respective window functions
  assuming $\Omega_{\rm HI}b_{\rm HI} = 0.43 \cdot 10^{-3}$.}
\label{3dPSGBTXWIG_IC}
\end{figure*}

The cross-power spectra of the intensity maps with the WiggleZ galaxy
survey for both fields is shown in Fig.~\ref{3dPSGBTXWIG_IC}, for a
range of different numbers of ICs. The errors in this figure are given
by the standard deviation of the estimates between the sub-datasets,
and the empty symbols mark negative correlations.  The cross-power
spectra converge with increasing number of ICs for both fields,
verifying that \textsc{fastica} does not subtract \tcm signal from
the data.

\begin{figure*}
\subfigure[10 ICs; 15hr-field]{
\includegraphics[width=0.45\textwidth]
{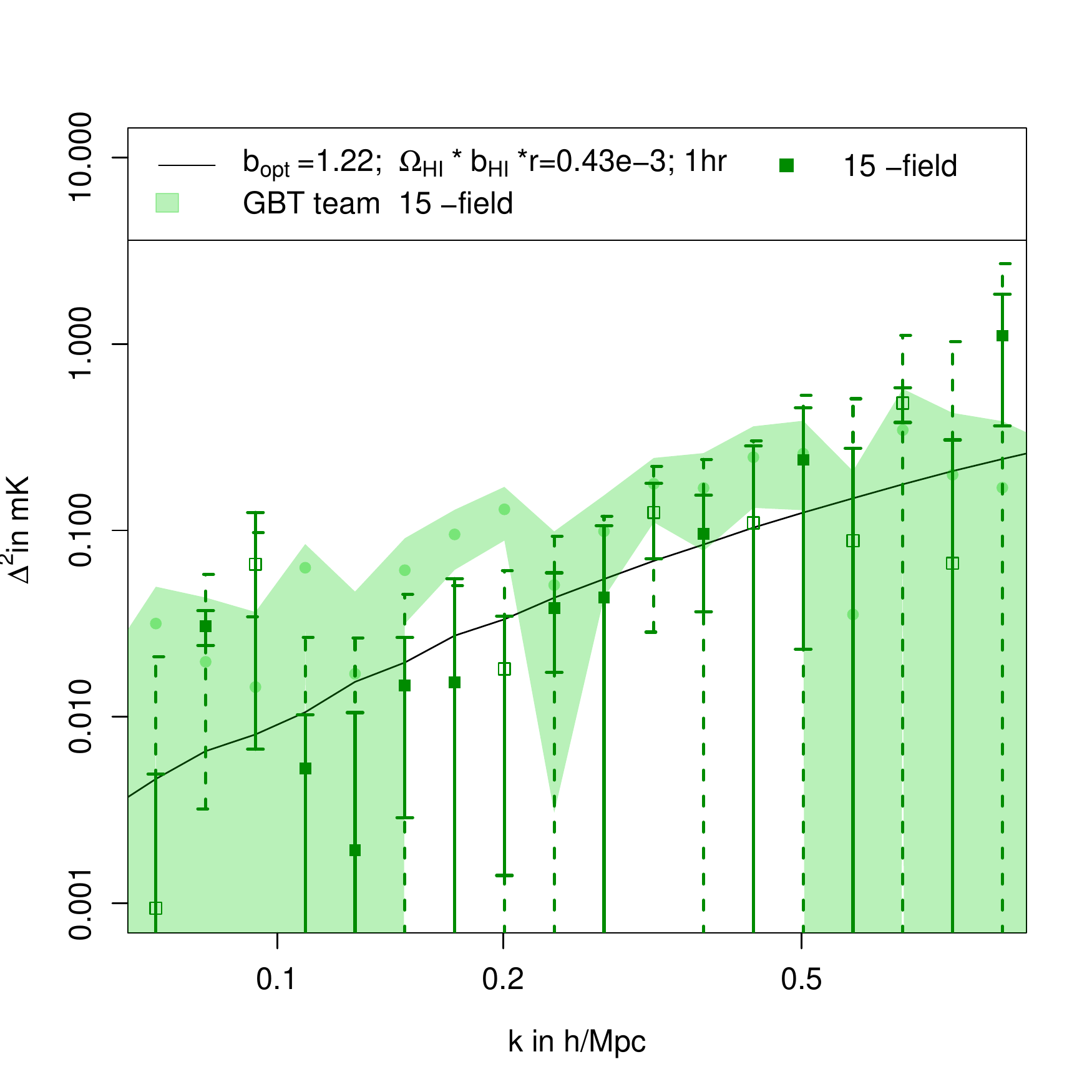}
\label{3dPSGBTXWIGa}
}\subfigure[20 ICs; 1hr-field]{
\includegraphics[width=0.45\textwidth]
{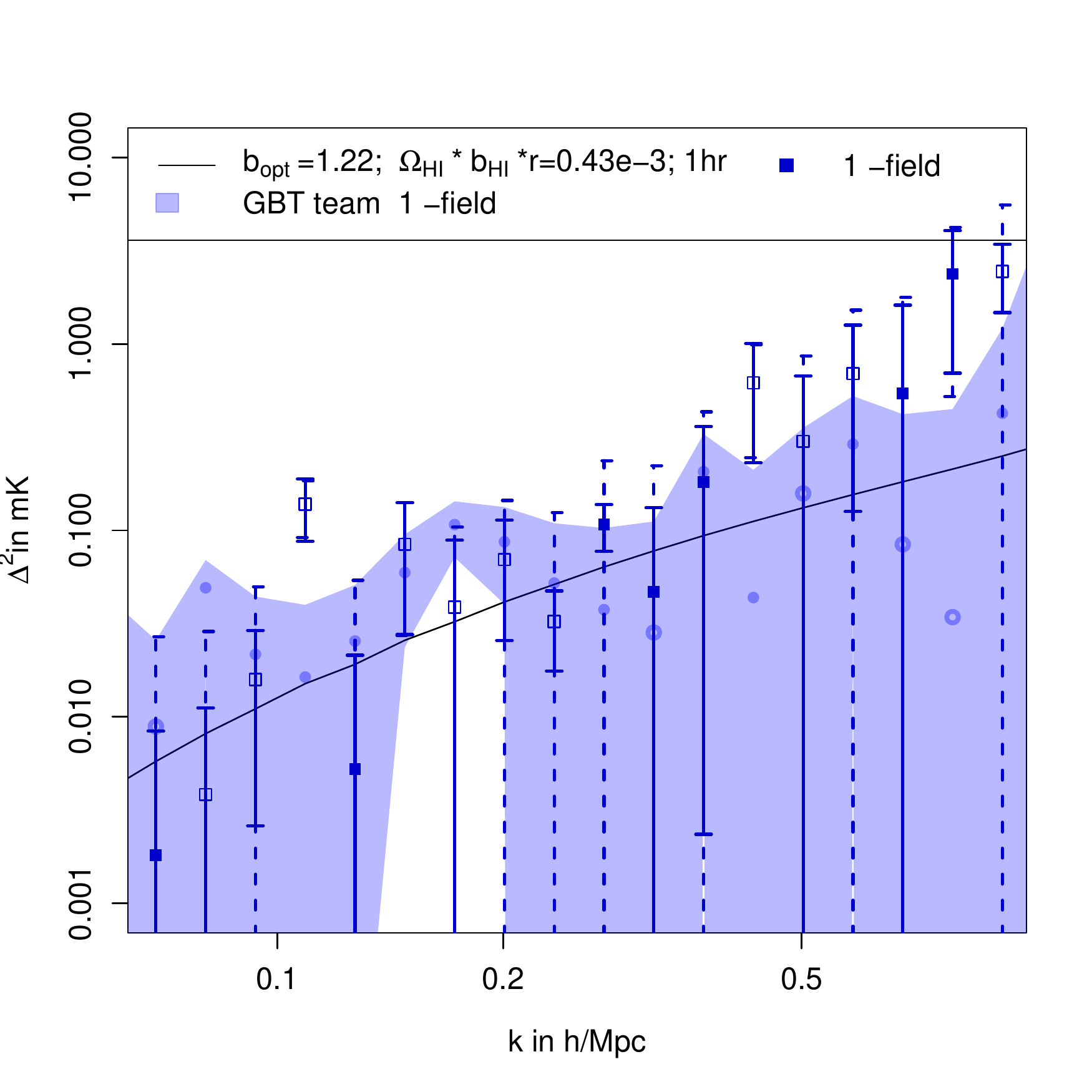}
\label{3dPSGBTXWIGb}
}
\caption[]{The 3D cross-power spectrum of the GBT intensity maps and
  WiggleZ galaxies, for the 15hr field (using foreground subtraction
  with 10 ICs) in the left-hand panel, and for the 1hr field (using 20
  ICs) in the right-hand panel.  The results of MA13 are marked by the
  shaded areas. The black lines are the theoretical model convolved
  with the respective window functions assuming $\Omega_{\rm HI}b_{\rm
    HI} = 0.43\cdot 10^{-3}$. The solid error bars are given by the standard deviation between sub-datasets and the dashed error bars using Equ.~\ref{Xnoise}.}
\label{3dPSGBTXWIG}
\end{figure*}

In Fig.~\ref{3dPSGBTXWIG} we show the cross-correlation for both the
15hr and 1hr fields, using 10 and 20 ICs respectively, in comparison
with the results of MA13, which are marked with blue and green shaded
areas. {Two measurement errors are shown: the standard
  deviation of the estimates between the sub-datasets (as the solid
  lines) and the theoretical expectation computed using
  Equ.~\ref{Xnoise} (as the dashed lines), which respectively provide
  an upper and lower limit of the measurement errors.}  Negative
cross-correlations are again indicated by empty symbols.
Fig.\ref{3dPSGBTXWIGa} demonstrates that our estimates generally agree
with the previous findings.

\section{Conclusions}
\label{conc}

In this study we present a thorough analysis of two intensity-mapping
fields observed by the GBT, previously analysed by
\cite{2013ApJ...763L..20M} and \cite{2013MNRAS.434L..46S}.  Our
pipeline includes a Fourier-based, weighted power spectrum estimator
for auto-correlations and cross-correlations with galaxy surveys.  We
remove the diffuse Galactic foregrounds and point source contamination
with \textsc{fastica}, which separates components based on a measure
of their non-Gaussianity.  The subtraction fidelity and systematic
errors are investigated for analyses with different numbers of ICs,
showing that the residual maps converge and the results are not
dependent on this choice.  We explore different masking of the maps to
reduce strong noise contamination at the edges of the fields.  We
confirm that \textsc{fastica} is well-suited for subtracting the
Galactic and non-Galactic foregrounds from intensity mapping data
since, by construction, it does not remove {Gaussian \tcm ~signal but can not prevent from removing the possibly non-Gaussian \tcm~signal}.

The auto-correlation of the residual intensity maps from
\textsc{fastica} has a higher amplitude than the previous measurements
by MA13 and SW13.  This is because \textsc{fastica} is a conservative
foreground removal technique compared to the SVD method.  Both
techniques measure auto-correlation power significantly above our
current best guess of the cosmological signal, indicating severe
systematic contamination in the current datasets.  The cross-power
spectrum between the intensity map and the WiggleZ galaxy survey
converges with increasing number of ICs, and is in reasonable
agreement with the measurements of MA13.

{We conclude that SVD and \textsc{fastica} serve as
  complementary tools for exploring the systematics and quality of
  foreground removal in noise-dominated intensity mapping datasets.}
In future work, we are planning to combine both techniques in order to
exploit their individual advantages in the data reduction.

\section*{Acknowledgments}

We thank the anonymous referee for their useful comments and suggestions.
Parts of this research were conducted by the Australian Research
Council Centre of Excellence for All-sky Astrophysics (CAASTRO),
through project number CE110001020.  CB acknowledges the support of
the Australian Research Council through the award of a Future
Fellowship.

\bibliographystyle{mn2e}
\bibliography{bib} 

\label{lastpage}

\end{document}